\documentclass[
 reprint,
longbibliography ,
 aps,
  twoside,
 superscriptaddress,
bibnotes
]{revtex4-1}

\usepackage{physics}
\usepackage{dsfont}
\usepackage{color,newfloat}
\usepackage{graphicx}
\usepackage{dcolumn}
\usepackage{bm}
\usepackage{amsmath,amssymb,amsthm}
\usepackage{mathtools}
\usepackage{multirow}
\usepackage{hhline, booktabs}
\usepackage{rotating}

\newcommand{\I}{\mathrm{i}}
\newtheorem{formula}{Formula}

\usepackage{chngcntr} 

\usepackage{enumerate}

\usepackage{algorithm}
\usepackage[noend]{algpseudocode}
\usepackage{etoolbox}

\begin{document}


\title{
A scheme for universal high-dimensional quantum computation with linear optics
}

\author{Stefano Paesani}
\email{stefano.paesani@nbi.ku.dk}
\affiliation{Quantum Engineering Technology Labs, H. H. Wills Physics Laboratory and Department of Electrical and Electronic Engineering, University of Bristol, Bristol BS8 1FD, UK}
\affiliation{Center for Hybrid Quantum Networks (Hy-Q), Niels Bohr Institute, University of Copenhagen, Blegdamsvej 17, DK-2100 Copenhagen, Denmark}
\author{Jacob F. F. Bulmer}
\affiliation{Quantum Engineering Technology Labs, H. H. Wills Physics Laboratory and Department of Electrical and Electronic Engineering, University of Bristol, Bristol BS8 1FD, UK}
\author{Alex E. Jones}
\affiliation{Quantum Engineering Technology Labs, H. H. Wills Physics Laboratory and Department of Electrical and Electronic Engineering, University of Bristol, Bristol BS8 1FD, UK}
\author{Raffaele Santagati}
\affiliation{Quantum Engineering Technology Labs, H. H. Wills Physics Laboratory and Department of Electrical and Electronic Engineering, University of Bristol, Bristol BS8 1FD, UK}
\affiliation{International Iberian Nanotechnology Laboratory (INL), Av. Mestre Jos\'{e} Veiga 4715-330 Braga, Portugal}

\author{Anthony Laing}

\affiliation{Quantum Engineering Technology Labs, H. H. Wills Physics Laboratory and Department of Electrical and Electronic Engineering, University of Bristol, Bristol BS8 1FD, UK}


\begin{abstract}
Photons are natural carriers of high-dimensional quantum information,
and, in principle, can benefit from
higher quantum information capacity
and noise-resilience. 
However,
schemes to generate the resources required
for high-dimensional quantum computing
have so far been lacking in linear optics.
Here, we show how to generate GHZ states in arbitrary dimensions and numbers of photons using linear optical circuits described by Fourier transform matrices.
Combining our results with recent schemes for qudit Bell measurements, we show that  universal linear optical quantum computing can be performed in arbitrary dimensions.
\end{abstract}

\date{\today}

\maketitle



Photonics is a sophisticated platform
for the development of quantum technologies,
from quantum processors
to distributed quantum communication~\cite{Kok2007, obrien2009, yin2017, Wang2019,rudolph2017}.
Until now, linear optical architectures have focused on encoding photons as qubits (two-level systems). 
Yet using higher dimensional systems -- qudits -- can in principle improve the information capacity
and noise tolerance of computational resources,
and potentially unlock new routes to fault-tolerant quantum computing and distributed quantum networks~\cite{Erhard2020}.
Qudits can naturally be encoded in photons using $d$ orthogonal optical modes in a variety of degrees of freedom, e.g. spatial modes~\cite{wang2018,vigliar2020,gomez2020}, orbital angular-momentum~\cite{dada2011,fickler2012,malik2016,Luo2019,erhard2018}, optical frequencies~\cite{kues2017,reimer2019}, and time-bins~\cite{Lee2016, Martin2017}. 
High-precision control and arbitrary operations of single qudits have been demonstrated using programmable interferometers~\cite{carolan2015,wang2018,carine2020,vigliar2020}.
However, architectures for universal quantum photonic processors based on higher-dimensional systems have, so far, been absent.
While there has been significant experimental progress in photonic qudit entanglement generation~\cite{wang2018,vigliar2020,gomez2020,dada2011,fickler2012,malik2016,kues2017,reimer2019,Martin2017,Luo2019,erhard2018},
the post-selected schemes used so far can only generate a limited set of high-dimensional entangled states and present no clear route to scalability~\cite{gu2019,krenn2020}.
In fact, in contrast to the qubit case~\cite{knill2001,Browne2005}, even  determining which high-dimensional entangled states can be generated with single photons and linear optics has so far remained an open problem~\cite{Erhard2020}. 

Here, we answer this question by showing that all high-dimensional entangled states with fixed numbers of photons can be generated with linear optics and, in principle, with a scalable architecture.
That is, we show that universal linear optical quantum computing (LOQC) is possible in arbitrary dimensions. 
Key to this result are linear optical schemes for the generation of heralded $N$-photon GHZ entanglement in arbitrary dimension $d$. 
Combining these schemes with previous results on Bell measurements with photonic qudits to fuse $d$-dimensional GHZ states, and techniques from qubit LOQC architectures, we obtain a scheme for universal measurement-based quantum computing with photonic qudits.  

 \begin{figure}[]
  \centering
  \includegraphics[
  trim=0 0 0 -10,
  width=0.5 \textwidth]{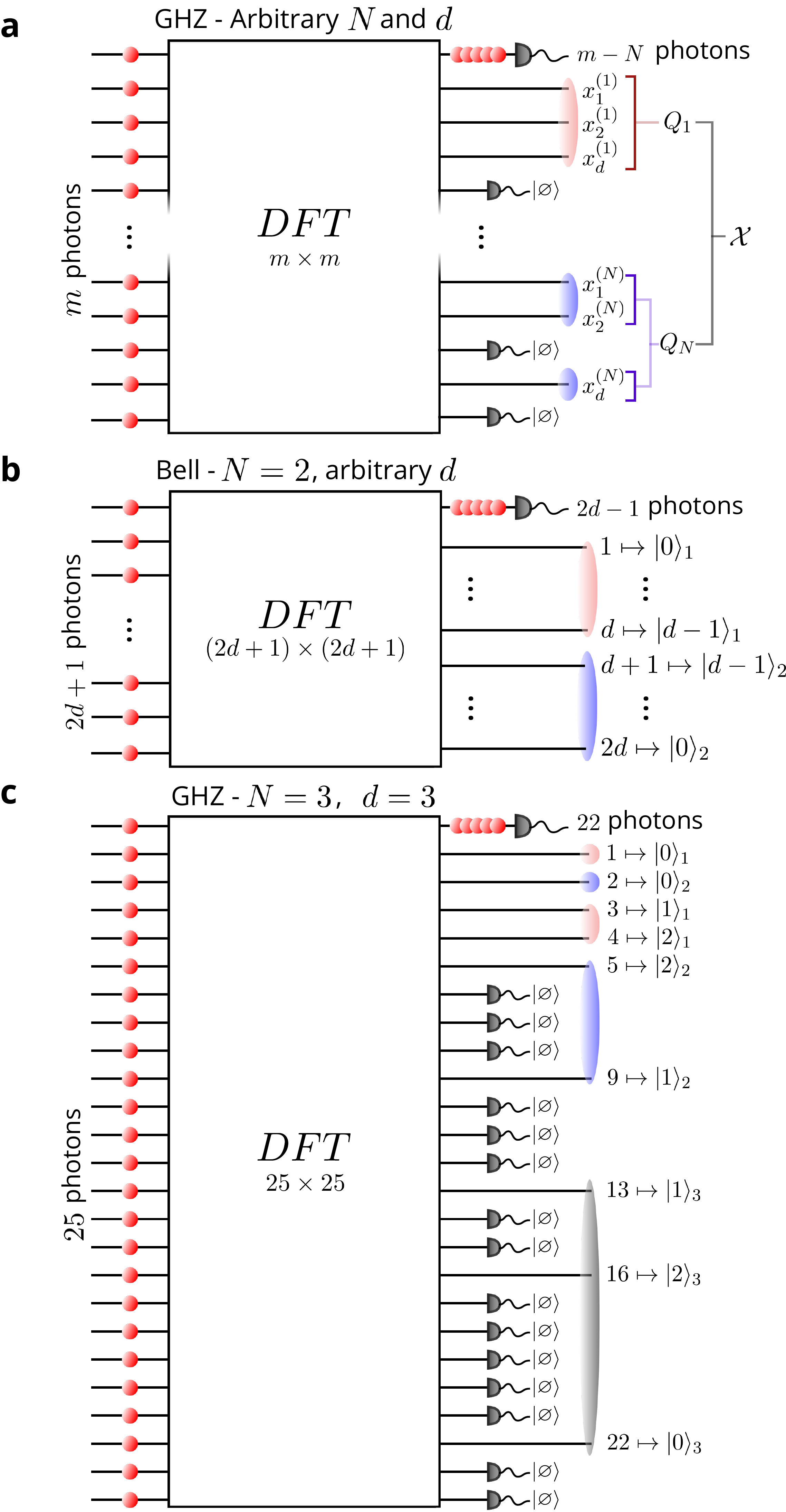}
  \caption{
    \textbf{a.} General schematic for the heralded generation of $N$-photon $d$-dimensional GHZ states via a DFT interferometer. The scheme requires a single number resolving detector for heralding in the zero-th output mode, while the rest are threshold detectors heralding the vacuum.
    \textbf{b.} Optimised solution for the heralded generation of Bell states ($N=2$) in arbitrary dimension using the DFT scheme, requiring $m=2d+1$. The correspondence between the optimal modes and the associated computational states of each qudit is shown.
    \textbf{c.} Optimised solution for the heralded generation of 3-photon 3-dimensional GHZ states using the DFT scheme, requiring $m=25$. Output modes associated to the three different qudits are highlighted with different colours, and the associated qudit computational states are labelled.
  } 
  \label{Fig_GHZ}
\end{figure}

\noindent \textbf{Heralded high-dimensional GHZ entanglement.} 
The GHZ states for $N$ photons, each encoding a qudit of dimension $d$, are defined as
\begin{equation}
    \ket{\mathrm{GHZ}(N,d)} = \frac{1}{\sqrt{d}} \sum_{k=0}^{d-1} \ket{k}^{\otimes N},
    \label{ghz_def}
\end{equation}
and represent the initial resource states
in an architecture that builds cluster states of qudits
for universal quantum computing.
Our scheme to generate these states,
shown schematically in  Fig.~\href{Fig_GHZ}{\ref{Fig_GHZ}a},
consists of single photon sources, linear optical elements, and photon-number resolving detectors.
Our scheme exploits the zero-transmission law (ZTL) in discrete Fourier transform (DFT) interferometers (previously investigated e.g. for the verification of boson sampling protocols~\cite{tichy2014,carolan2015,crespi2016}).
In particular, we use a result of Tichy et al.~\cite{Tichy2010} which states that, if $m$ single photons are sent in the $m$ individual input modes of an $m$-mode DFT, described by the unitary $U_{j,k} = \exp(jk \frac{2\pi \I}{m})/\sqrt{m}$, then the output configurations $\vec{c}$ with non-zero amplitude must satisfy 
\begin{equation}
\sum_{i=1}^m c_i \mod{m}  = 0, 
\label{eq_ZTL}
\end{equation}
where $c_i$ represents each particle’s output port, with modes indexed from $0$ to $m-1$. 
For example, $\vec{c}=(0,1,1,4)$ represents the configuration with one photon in each of modes $0$ and $4$, two photons in mode $1$, and zero photons in all remaining modes. 
The ZTL condition is valid for any value of $m$. 
To describe how the ZTL can be used to generate GHZ states, we introduce a notation for the encoding of individual qudits in the output modes of the DFT.
Each qudit of dimension $d$ is encoded in a single photon propagating through $d$ optical modes (see e.g.~Refs.\cite{wang2018,vigliar2020}).
We will denote as $\mathcal{X} \subset \{0,\ldots,m-1\}$ the total set of $Nd$ modes used to encode the $N$ qudits, while its complement $\bar{\mathcal{X}}=\{0,\ldots,m-1\} \setminus \mathcal{X}$ represents all the auxiliary modes used for heralding
the desired state. 
Each of the $N$ qudits is defined via a set of $d$ distinct modes $Q_i = \{x^{(i)}_{0}, x^{(i)}_{1}, \ldots, x^{(i)}_{d-1}\} \subset \mathcal{X}$, with $x^{(i)}_{k} \neq x^{(i)}_{k'}$ if $k \neq k'$, so that all the qudits $\{Q_i\}$ form a partition of $\mathcal{X}$. 
We say that the $i$-th qudit is in the logical state $\ket{k}_i$ if a single photon is present in mode $x^{(i)}_{k}$, while all other modes in the  set $Q_i$ are vacuum.
For simplicity, we will focus on heralding configurations where all the $m-N$ auxiliary photons are detected in the zero-th mode, and vacuum in the remaining heralding modes $\bar{\mathcal{X}}\setminus \{0\}$, as shown in Fig.~\href{Fig_GHZ}{\ref{Fig_GHZ}a}. 
In this way, all auxiliary photons give zero contributions to the sum in Eq.~\ref{eq_ZTL}, and only contributions from the $N$ encoding photons remain. 
Under this condition, we can write the set of all possible remaining $N$-photon output states induced by the heralding on $\bar{\mathcal{X}}$ and the ZTL as
\begin{align}
    \mathcal{B}_{\mathcal{X}}=\Big\{ \vec{b}_k = &( y_{k,1}, y_{k,2}, \ldots, y_{k,N} )\Big| \nonumber \\
    &\sum_{i=1}^N y_{k,i} \mod m = 0,\  y_{k,i} \in \mathcal{X}
    \Big\}.
\end{align}
\noindent Here $\vec{b}_k$ are the output configurations allowed by the ZTL, indexed by $k$, and $y_{k,i}$ represent each photon's output port for that configuration, as in Eq.~\ref{eq_ZTL}.  
We will now show that the following conditions for the set $\mathcal{B}_{\mathcal{X}}$ are sufficient (but not necessary) to obtain a GHZ state at the output:
\begin{enumerate}
    \item $\mathcal{B}_{\mathcal{X}}$ contains exactly $d$
    configurations, i.e. $|\mathcal{B}_{\mathcal{X}}|=d$. 
    \item $\mathcal{B}_{\mathcal{X}}$ forms a partition of $\mathcal{X}$.
    \item For all $\vec{b}_k \in \mathcal{B}_{\mathcal{X}}$,  $\sum_{i=1}^N y_{k,i} = m$. This is a slightly more restrictive form of the ZTL.
\end{enumerate}
The first step is to show that, if these three conditions are satisfied, we can specify $N$ well-defined photonic qudits, given by $Q_i = \{x^{(i)}_k = y_{k,i}\}_{k=0,\ldots,d-1}$. 
Note that, because the sets $\{Q_i\}$ are given by a simple transposition of indices of  $\mathcal{B}_{\mathcal{X}}$, condition 2 immediately implies that $\{Q_i\}$ also forms a partition of $\mathcal{X}$. 
Moreover, conditions 1 and 2 imply that all $y_{k,i}$ are different, i.e. $y_{k,i}\neq y_{k',i'}$ if $k \neq k'$ or $i\neq i'$. 
In fact,  because $\mathcal{B}_{\mathcal{X}}$ contains exactly $d$ configurations of $N$ elements, if any two $y_{k,i}$ were identical then necessarily $\left| \bigcup_k \vec{b}_k \right|< Nd$, and thus $\mathcal{B}_{\mathcal{X}}$ could not form a partition of $\mathcal{X}$, given that $\left|\mathcal{X} \right| = Nd$. 
Finally, from their definition and the fact that the elements $y_{k,i}$ are all different, it immediately follows that each set $Q_i$ contains exactly one output photon.
Our specification for the sets $\{Q_i\}$ therefore defines $N$ disjoint sets of $d$ different modes, each set containing exactly one photon, and thus provides a valid encoding for the $N$ qudits. 
We are now left to verify that the state of the $N$ qudits is in fact a GHZ state.   
Note that, with the definition used for the $Q_i$s, the $d$ elements $\vec{b}_k$ of $\mathcal{B}_{\mathcal{X}}$ correspond to the logical $N$-qudit states $\vec{b}_k \mapsto \ket{k,k,\ldots,k}$, $k\in{0,\ldots, d-1}$. 
Due to the ZTL, the total output state is therefore a superposition of the states $\ket{k,k,\ldots,k}$. 
As shown in Appendix~\ref{SupplInfo:GHZschemeCalculations}, condition 3 ensures that all amplitudes in such superposition are uniform and non-zero, thus providing the desired $N$-photon GHZ state in $d$ dimensions. 

\ \\
\noindent\textbf{General and optimised schemes.} 
The task of generating GHZ states for a given photon number $N$ and dimension $d$ can now be reduced to a combinatorial number theory problem: finding an integer number $m$ and a set $\mathcal{X}\subseteq \{1,2,\ldots,m-1\}$ ($0$ is occupied for the heralding) so that $\mathcal{B}_{\mathcal{X}}$  satisfies conditions 1-3. Solutions can be found for any $N$ and $d$. 
An example of such general solutions is given by the set
\begin{equation}
    {\mathcal{X}} = \{ N^{i-1}\}_{i \in [1,(N-1)d]}
    \cup \{m - \sum_{j=0}^{N-2} N^{jd+k}\}_{k \in [0, d-1]},
\end{equation}
with $m=(N^{Nd} - 1)/(N^d - 1)$, with details in Appendix~\ref{SupplInfo:GHZExamples}. 
While this particular solution is highly sub-optimal in the number of resources and success probability, and likely not suitable for practical implementations, it is general and shows that arbitrary $N$-photon $d$-dimensional GHZ states can, in principle, be generated with linear optics. 
More efficient solutions can be found on a case-by-case basis. 
For example, in the simple case with $N=2$, i.e. heralded generation of qudit Bell pairs, a solution for any $d$ can easily be found by taking $m=2d+1$ and $\mathcal{X}=\{1,2,\ldots,2d\}$. This gives  $\mathcal{B}_{\mathcal{X}}=\{ (1,2d),(2,2d-1),\ldots,(d,d+1)\}$
and qudit modes $Q_1=\{1,2\ldots, d\}$, $Q_2=\{2d,2d-1\ldots, d+1\}$, as shown in Fig.~\href{Fig_GHZ}{\ref{Fig_GHZ}b}. 
The success probability for the state generation is given by $d(2d-1)!/(2d+1)^{2d-1}$ (see Appendix~\ref{SupplInfo:GHZschemeCalculations}, and Appendix~\ref{bell_states} for additional schemes for qudit Bell states). 

 \begin{figure*}[]
  \centering
  \includegraphics[
  trim=0 0 0 -10,
  width=0.9 \textwidth]{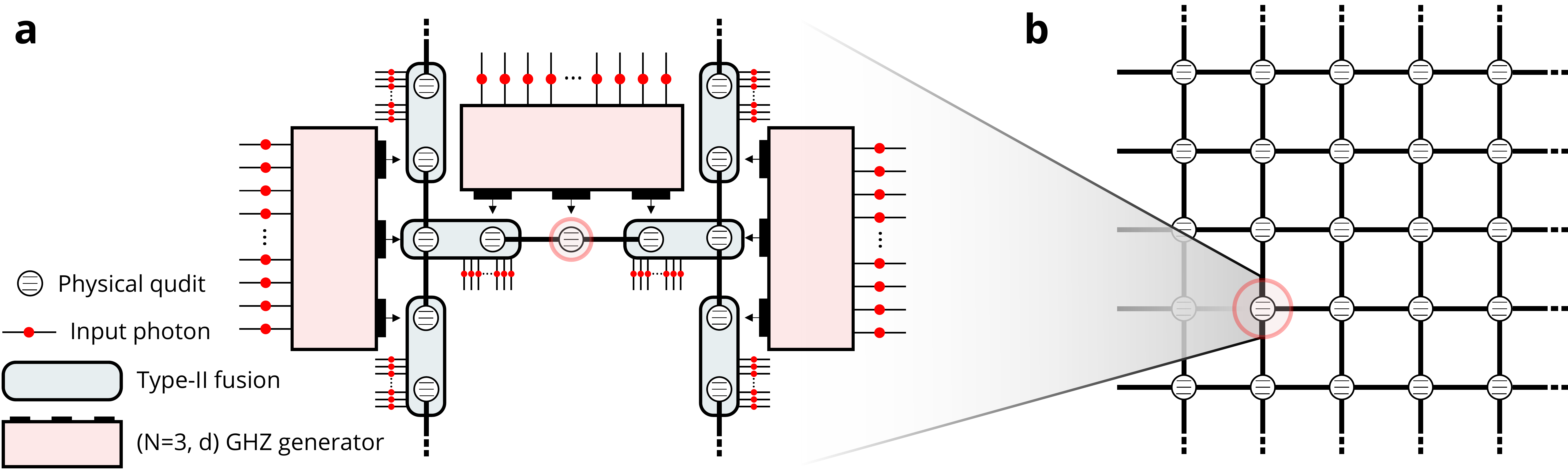}
  \caption{
    Example of modular architecture for constructing universal cluster states from three-photon GHZ states with linear optics.
    Each module, shown in \textbf{a}, arranges three-photon GHZ state generators and fusion gates such that a single $d$-dimensional qudit is linked to the four neighbouring qudits in the universal square lattice in \textbf{b}. 
  } 
  \label{Fig_LOQC}
\end{figure*} 

For small values of $N$ and $d$, optimised solutions for our heralded GHZ generation scheme can be found numerically. 
For example, in Fig.~\href{Fig_GHZ}{\ref{Fig_GHZ}c} we report the optimised solution for the case of heralded GHZ generation for $(N,d)=(3,3)$, which requires $m=25$.
The list of modes used for the encoding in this case is $\mathcal{X}= \{1,2,3,4,5,9,13,16,22\}$, and the only three-element combinations (including repetitions) that sum up to multiples of $m$, given by $\mathcal{B}_{\mathcal{X}}=\{(1,2,22),(3,9,13), (4,5,16) \}$, satisfy conditions 1-3. 
This therefore provides heralded GHZ generation for the three qudits defined in the modes $Q_1=\{1,3,4\}$, $Q_2=\{2,9,5\}$, $Q_3=\{22,13,16\}$. 
Note that, at the output of the DFT, adjacent modes associated to different qudits can be interleaved; in such cases, a network of swaps is required to separate the different qudits and address them individually.
However, without further optimisation, the three-dimensional GHZ generation success probability remains very small, approximately $10^{-10}$. %

The main reason for low success probabilities is the use of a single heralding pattern amongst exponentially many possible outcomes. 
While the choice of a single heralding pattern was done to simplify the treatment in the general case of arbitrary $N$ and $d$, for a given $N$ and $d$ many more heralding patterns are likely to generate GHZ entanglement.
We show in Appendix~\ref{SupplInfo:ProbBoost} that these (combinatorially many) valid heralding patterns can be used, in conjunction with feed-forward operations and balancing circuitry, to significantly improve the success probability.
For example, 
Monte Carlo simulations of the $(N,d)=(3,3)$ GHZ generation scheme of Fig.~\href{Fig_GHZ}{\ref{Fig_GHZ}c} show that the success probability is boosted to approximately $10^{-4}$.
This indicates that many orders of magnitude improvements can be found through solution-specific optimisations.
We estimate that, with this optimisation, heralded $(3,3)$ GHZ generation could be achieved at $0.7$~kHz rates with state-of-the-art quantum photonic hardware. Furthermore, in Appendix 2 we report an algorithm to estimate optimized solutions for larger values of $N$ and $d$.
%


\ \\
\noindent\textbf{Constructing universal cluster states of qudits.}
Measurement-based quantum computing (MBQC)~\cite{Raussendorf2001}
in linear optics
typically proceeds by connecting small entangled resource states using probabilistic fusion gates~\cite{Browne2005}, to build large cluster states.
Similarly, high-dimensional photonic GHZ states can be used as building-blocks to construct large high-dimensional cluster states for MBQC.
Two recent protocols for type-II fusion (destructive Bell state measurements) of arbitrary dimensional qudits have been proposed independently:
Luo et al.~\cite{Luo2019}
use $d - 2$ unentangled ancillary single photons; Zhang et al.~\cite{Zhang2019}
use entangled $d-2$ Bell states. 
The success probability of both fusion gates scales $\approx 1/d^2$.
As shown in Fig.~\ref{Fig_LOQC}, we can combine type-II high-dimensional fusion operations with three-photon high-dimensional GHZ generators in a modular approach to build up universal cluster states of qudits.
The square lattice of qudits shown in  Fig.~\href{Fig_LOQC}{\ref{Fig_LOQC}b} is an example of a cluster state universal for high-dimensional MBQC~\cite{Zhou2003,hall2005}.
Fig.~\href{Fig_LOQC}{\ref{Fig_LOQC}a} shows a module for the architecture, where multiple GHZ states are fused together to link a single computational photonic qudit to the rest of the lattice. 
Once such states are built, universal measurement-based high-dimensional quantum computing can be performed~\cite{Raussendorf2001,Zhou2003,hall2005}: operations on the logical qudits encoded in the rows of the lattice are performed via measurement and feed-forward, with the output state encoded in the qudits of the last layer of the lattice. 
Because the resource is universal, any pure high-dimensional multi-photon state can be prepared as a result of the computation up to arbitrary precision. 
This implies that the generation of arbitrary quantum states comprised of $d$-dimensional photons, for any $d$, is possible using only linear optics.
On the other hand, because both the GHZ generation and the type-II fusion gates have low success probability, the total probability to successfully generate high-dimensional qudit cluster states can in general vanish quite rapidly when increasing the number of qudits or the dimensionality. 
Nevertheless, by adapting techniques already developed for qubit-based loss-tolerant LOQC architectures, the approach can in principle be made scalable and the total success probability boosted to near-unity~\cite{Browne2005,Kieling2007, Gimeno2015, Li2015}.  
For example, because the GHZ generation is heralded, gate multiplexing can be used to render the production of GHZ states  near-deterministic with a resource overhead that scales approximately linearly with the generation success probability~\cite{Gimeno2015,gimeno2017}.
Repeat-until-success proposals provide a flexible approach to correct for the limited success probability of the fusion gates, at the cost of requiring quantum memories~\cite{Browne2005}. 
Moreover, if the fusion success probability is improved above the percolation threshold of the lattice used, ballistic architectures can be used to correct the unsuccessful fusion gates directly on the generated lattice without the use of quantum memories~\cite{Kieling2007, Gimeno2015}. 
However, this approach would likely require the use of lattices with increased valency to bring the percolation threshold above the current qudit fusion gates success rates~\cite{Galam1996}. 
In addition, it may be possible to increase the success probability of the qudit fusion gates by using additional ancilla resources, as has been shown for qubits~\cite{Grice2011, Ewert2014}.


 \begin{figure}[]
  \centering
  \includegraphics[
  trim=0 0 0 -10,
  width=0.475 \textwidth]{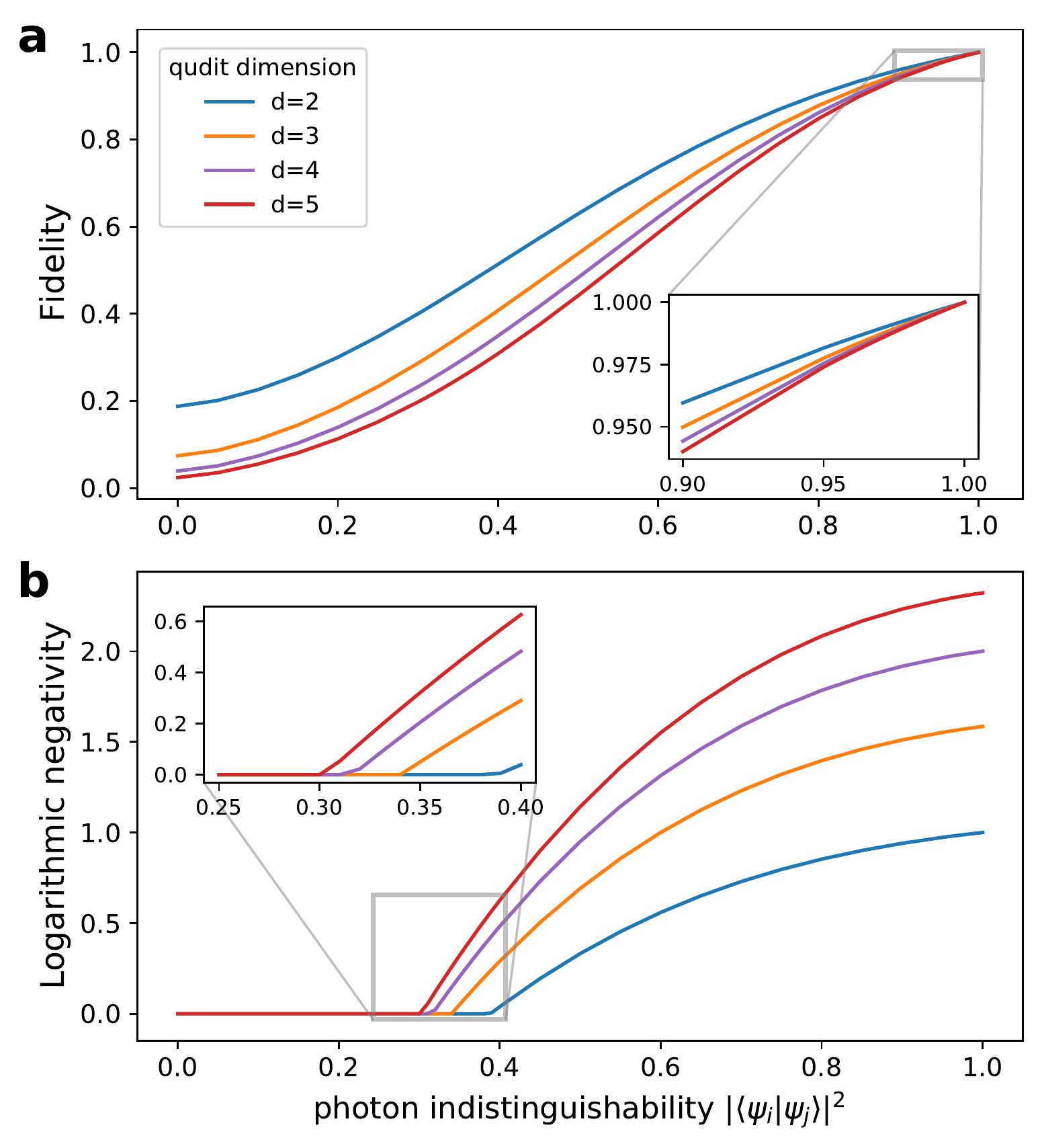}
  \caption{Robustness of the qudit Bell state generation circuits against photon indistinguishability. \textbf{a.} Quantum fidelity of the generated qudit Bell state with the ideal, plotted against photon indistinguishability. 
  \textbf{b.} The logarithmic negativity bound on the distillable entanglement from the generated state with photon indistinguishability.} 
  \label{Fig_DistPlot}
\end{figure} 

\ \\
\textbf{Robustness against photon distinguishability.} 
While requiring additional resources compared to qubit-based approaches, qudits can provide improved robustness to noise, with potential benefits for quantum communication and fault-tolerant quantum technologies~\cite{Sheridan2010,Campbell2014,Watson2015}.
An important source of noise in quantum photonics is distinguishability between optical modes, which can arise from imperfections in solid-state or spontaneous photon emitters. 
Here, we numerically analyse how distinguishability affects the multi-photon high-dimensional entanglement generated using the schemes proposed in Fig~\ref{Fig_GHZ}. 
%
We assume that all photons have pure internal states such that they all share the same value of pairwise indistinguishability $\vert\langle \psi_{i}\vert \psi_{j}\rangle\vert^{2}$. Thus all pairs of photons would exhibit the same Hong-Ou-Mandel interference visibility.
%
%
To determine the effect of distinguishability, we numerically reconstruct the simulated output heralded state within the qudit space for different values of indistinguishability $\vert\langle\psi_{i}\vert\psi_{j}\rangle\vert^2$. 
The noisy state generation is simulated using an approach by Tichy describing multi-photon interference of partially distinguishable photons~\cite{Tichy2015} (see Appendix~\ref{SupplInfo:DistAnalysis} for details). 
Because the number of photons involved in the scheme increases rapidly with $N$ and $d$, this simulation quickly becomes intractable when increasing the heralded state complexity. 
Nevertheless, for the scheme shown in Fig.~\href{Fig_GHZ}{\ref{Fig_GHZ}b} with $N=2$ and low values of $d$, the simulation remains tractable on a standard laptop. 
For various levels of distinguishability, we show in Fig.~\ref{Fig_DistPlot}a the fidelity of the generated state to the ideal qudit Bell state, and in Fig.~\ref{Fig_DistPlot}b the logarithmic negativity as a figure of merit to assess the generated entanglement. 
The logarithmic negativity is an entanglement monotone that upper bounds the distillable entanglement, which in turn quantifies the amount of pure state entanglement that can be extracted  under local operations and classical communication~\cite{Vidal2002,Plenio2005}.
Distinguishability weakens the interference governing the ZTL, and therefore, as expected, both the state fidelity and generated entanglement decrease when the indistinguishability is reduced.
However, while the fidelity is always lower for higher dimensional states, the negativity reaches zero at larger values of photon distinguishability for higher dimensions, as highlighted in the inset of Fig.~\ref{Fig_DistPlot}b. 
This indicates that for increased dimensionality, the generated entanglement can endure higher levels of photon distinguishability, even if the number of imperfect input photons is larger.

\ \\
\noindent\textbf{Discussion.}
%
%
The large (but polynomial) resource overheads
of high-dimensional LOQC
could be compensated to some extent
by quantum error correction protocols that are more efficient
due to robustness to noise that increases with qudit dimension.
Furthermore, and similarly to the improvements made in efficiency for qubit-based LOQC~\cite{knill2001,Browne2005,Gimeno2015}, we expect our results to provide a first and important step for developing high-dimensional LOQC architectures closer to mid-term technological capabilities. 
For example, as detailed in Appendix~\ref{SupplInfo:ProbBoost}, by considering all valid heralding events in our GHZ generation scheme for the particular case with $d=3$ (see Fig.~\href{Fig_GHZ}{\ref{Fig_GHZ}c}), the resource overheads for multiplexing are reduced from $\mathcal{O}(10^{10})$ to $\mathcal{O}(10^{4})$, showing that enormous improvements are possible via solution-specific optimisation. 
Crucially, this work proves that such solutions can exist for any dimension.
%



\bibliographystyle{apsrev4-1}
\bibliography{refs.bib}

\begin{footnotesize}
\noindent
\ \\

\textbf{Acknowledgements.}
\noindent We thank C. Vigliar, D. Bacco, M. Erhard and M. Krenn for useful discussions. We acknowledge support from the Engineering and Physical Sciences Research Council (EPSRC) Hub in Quantum Computing and Simulation (EP/T001062/1) and the Networked Quantum Information Technologies (EP/N509711/1). Fellowship  support  from  EPSRC  is acknowledged by A.L. (EP/N003470/1).

\end{footnotesize}


\newpage 
\clearpage

\pagenumbering{arabic}

\onecolumngrid

\appendix

\setcounter{page}{1}

\renewcommand{\thesection}{\arabic{section}}

\renewcommand{\thefigure}{\textbf{S\arabic{figure}}}
\renewcommand{\figurename}{\textbf{Supplementary Figure}}
\setcounter{figure}{0} 

\renewcommand{\thetable}{\textbf{S\arabic{table}}}
\renewcommand{\tablename}{\textbf{Supplementary Table}}
\setcounter{table}{0}

\counterwithout{equation}{section}
\renewcommand{\theequation}{S\arabic{equation}}
\setcounter{equation}{0}


\section{Calculation of output amplitudes and success probabilities in the $N$-photon $d$-dimensional heralded GHZ generator}
\label{SupplInfo:GHZschemeCalculations}

Here we provide the technical details for the derivation of the output amplitudes in the DFT-based $N$-photon $d$-dimensional heralded GHZ generator described in the main text. We start by reporting some useful formulas for the analytical calculation of bosonic evolutions in DFT interferometers, and then use these results to obtain the output amplitudes and success probabilities of the heralded GHZ generator circuit.

\subsection{Some useful formulas for bosonic transformations in Discrete Fourier Transforms}

\subsubsection{Definitions}

\begin{itemize}
    \item $P_{m}$: set of all permutation of $\{0,\ldots,m-1\}$.
    \item $P_{m}\setminus b$: set of all permutation of $\{0,\ldots,m-1\}\setminus b$, which is the set of all numbers between 0 and $n-1$ excluding $b\in \{0,\ldots,m-1\}$.    
\end{itemize}

\subsubsection{Formulas}

\begin{formula} \textbf{Zero-Transmission Law (ZTL)}~\cite{Tichy2010}: Given a vector $\vec{a}=(a_0, a_1, \ldots, a_{m-1})$, with $a_i\in \mathbb{N}_0$:  
\begin{equation*}
  \text{if}\qquad \sum_{i=0}^{m-1} a_i \neq 0 \mod m, \qquad \text{then} \qquad \sum_{\sigma \in P_{m}}\exp\left[\left(\sum_{i=0}^{m-1} \sigma_i a_i \right) \frac{2\pi \I}{m} \right] = 0.
\end{equation*}
\label{Formula:ZTL}
\end{formula}

\begin{proof}
 Derived in Ref.~\cite{Tichy2010}.
\end{proof}

\begin{formula} Given a vector $\vec{a}=(a_0, a_1, \ldots, a_{n-1})$, with $a_i\in \mathbb{N}_0$, and $n\leq m$: 
\label{Formula:ZTLrestricted}
\begin{equation*}
  \text{if}\qquad \sum_{i=0}^{n-1} a_i \neq 0 \mod m, \qquad \text{then} \qquad \sum_{\sigma \in P_{m}}\exp\left[\left(\sum_{i=0}^{n-1} \sigma_i a_i \right) \frac{2\pi \I}{m} \right] = 0.
\end{equation*}
\end{formula}

\begin{proof}
 This corollary of the ZTL can be simply obtained by padding $m-n$ zeros to the vector $\vec{a}$, i.e. defining the $m$-element vector $\vec{b} =(a_0,\ldots,a_{n-1},0, \ldots,0)$. Obviously we have \begin{equation}
     \sum_{i=1}^{m-1} b_i = \sum_{i=1}^{n-1} a_i  \neq 0 \mod m, \quad \text{and} \quad \sum_{i=1}^{m-1} \sigma_i b_i = \sum_{i=1}^{n-1} \sigma_i a_i. 
 \end{equation}
We can then simply apply Formula~\ref{Formula:ZTL} to obtain
 
 \begin{equation}
    \sum_{\sigma \in P_{m}}\exp\left[\left(\sum_{i=0}^{n-1} \sigma_i a_i \right) \frac{2\pi \I}{m} \right] = \sum_{\sigma \in P_{m}}\exp\left[\left(\sum_{i=0}^{m-1} \sigma_i b_i \right) \frac{2\pi \I}{m} \right] = 0.
 \end{equation}
\end{proof}

\begin{formula} Given a vector $\vec{a}=(a_0, a_1, \ldots, a_{n-1})$, with $a_i\in \mathbb{N}_0$, such that $\sum_{i=0}^{n-1} a_i = m$ and $a_i<m \ \forall i$, we have 
\label{Formula:GeneralP}
\begin{equation*}
\sum_{\sigma \in P_{m}}\exp\left[\left(\sum_{i=0}^{n-1} \sigma_i a_i \right) \frac{2\pi \I}{m} \right] = (-1)^{n-1} m (n-1)! (m-n)!.
\end{equation*}
\end{formula}

\begin{proof}
Because $\sum_{i=0}^{n-1} a_i = m$ and $a_i<m$, then at least two elements in the vector are non-zero. Therefore, given also the symmetry under permutations of $\vec{a}$ of the sum in the formula, we can consider $a_0>0$ and $a_{n-1}>0$ without loss of generality. Defining 

 \begin{equation}
    T \coloneqq \sum_{\sigma \in P_{m}}\exp\left[\left(\sum_{i=0}^{n-1} \sigma_i a_i \right) \frac{2\pi \I}{m} \right],
 \end{equation}
 we can rewrite $T$ by fixing the $n$-th value of the permutation to be $b\in \{0,\ldots,m-1\}$ and summing over the permutations of the remaining $m-1$ elements and  over all possible values of $b$:
\begin{align}
    T \coloneqq \sum_{\sigma \in P_{m}}\exp\left[\left(\sum_{i=0}^{n-1} \sigma_i a_i \right) \frac{2\pi \I}{m} \right]  &= \sum_{b=0}^{m-1}  \exp\left( b a_{n-1} \frac{2\pi \I}{m} \right) \sum_{\sigma \in P_{m}\setminus b} \exp\left[\left(\sum_{i=0}^{n-2} \sigma_i a_i\right) \frac{2 \pi \I}{m} \right] \nonumber \\
    \label{eq:TrickComplete}
    &= \sum_{b=0}^{m-1}  \exp\left( b a_{n-1} \frac{2\pi \I}{m} \right) S_{n-1}(b)
\end{align}
where 
\begin{equation}
    S_{n}(b) \coloneqq  \sum_{\sigma \in P_{m}\setminus b} \exp\left[\left(\sum_{i=0}^{n-1} \sigma_i a_i\right) \frac{2 \pi \I}{m} \right].
\end{equation}
For all $n'<n$, we can expand $S_{n'}(b)$ in a convenient iterative form. The idea is to consider the sum for all permutations in $P_{m}$ excluding the permutations where $b$ actually appears in the first $n'$ elements. Given that only the first $n'$ elements are useful, the contributions in this sum are exactly the same as when summing over $P_{m}\setminus b$, but each one repeated $m-n'$ time (the possible ways to distribute $b$ in the remaining $m-n'$ elements). Considering also a symmetry over permutations of $(a_0, a_1, \ldots, a_{n-1})$, we can finally write

\begin{align}
    S_{n'}(b) &= \frac{1}{m-n'} \Bigg\{ \sum_{\sigma \in P_{m}} \exp\left[\left(\sum_{i=0}^{n'-1} \sigma_i a_i \right) \frac{2\pi \I}{m} \right]
    - n' \sum_{\sigma \in P_{m}\setminus b} \exp\left[\left(\sum_{i=0}^{n'-2} \sigma_i a_i  + b a_{n'} \right) \frac{2\pi \I}{m} \right]   \Bigg\} \\
    &= \frac{1}{m-n'} \Bigg\{ \sum_{\sigma \in P_{m}} \exp\left[\left(\sum_{i=0}^{n'-1} \sigma_i a_i \right) \frac{2\pi \I}{m} \right] - n'\exp\left(b a_{n'} \frac{2\pi \I}{m}\right)S_{n'-1}(b)\Bigg\}.
\end{align}
Because $\sum_{i=0}^{n-1} a_i = m$ and $a_{0}>0$ and $a_{n-1}>0$, for all $1 < n' < n$ we have
\begin{equation}
    \sum_{i=0}^{n'-1} a_i  \neq 0 \mod m, 
\end{equation}
which allows us to use Formula~\ref{Formula:ZTLrestricted} to set the first term in the expansion of $S_{n'}(b)$ to zero:

\begin{equation}
    \label{eq:rec_relations}
S_{n'}(b) = -\frac{n'}{m-n'} \exp\left(b a_{n'}\frac{2\pi \I}{m}\right)S_{n'-1}(b).
\end{equation}
The term $S_{0}(b)$ can be easily calculated as 
\begin{align}
S_0(b) &= \sum_{\sigma \in P_{m}\setminus b} \exp\left( \sigma_0 a_0 \frac{2\pi \I}{m} \right) = (m-1)! \sum_{x=0,\ x \neq b}^{m-1} \exp\left( x a_0 \frac{2\pi \I}{m} \right) \nonumber \\ 
&=(m-1)! \left[\sum_{x=0}^{m-1} \exp\left( x a_0 \frac{2\pi \I}{m} \right) - \exp\left( b a_0 \frac{2\pi \I}{m} \right) \right] \nonumber \\
&=(m-1)! \left[\frac{1- \exp\left(m a_0 \frac{2\pi \I}{m}\right)}{1- \exp\left(a_0 \frac{2\pi \I}{m}\right)} - \exp\left( b a_0 \frac{2\pi \I}{m} \right) \right] \nonumber \\
&= - (m-1)! \exp\left( b a_0 \frac{2\pi \I}{m} \right),
\end{align}
where we used that, because $0<a_0/m<1$, necessarily $1- \exp\left(2\pi \I a_0/m\right) \neq 0$.
Substituting in Eq.~\ref{eq:rec_relations} we obtain a simple recursive relation which provides
\begin{equation}
    S_{n'}(b)= (-1)^{n'} n'! (m-n'-1)! \exp\left[\left(\sum_{i=0}^{n'-2} \sigma_i a_i \right) \frac{2\pi \I}{m} \right].
\end{equation}
Substituting in Eq.~\ref{eq:TrickComplete} we finally obtain 
\begin{align}
    T &= (-1)^{n-1} (n-1)! (m-n)!  \sum_{b=0}^{m-1}  \exp\left[\left(\sum_{i=0}^{n-1} a_i \right) b \frac{2\pi \I}{m} \right] \\
    &= (-1)^{n-1} (n-1)! (m-n)! \sum_{b=0}^{m-1}  \exp\left(m b \frac{2\pi \I}{m} \right)\\
    &= (-1)^{n-1} m (n-1)! (m-n)!
\end{align}
which is interestingly independent on the vector $\vec{a}$.

\end{proof}

\subsection{Output amplitudes and success probabilities in the DFT-based GHZ generator}

An $n$-photon output configuration can be described by an $n$-element vector $\vec{x}=(x_0,x_1,\ldots,x_{n-1})$, with $x_i\in{0,\ldots,m-1}$, representing the $n$ output modes of the photons. Each configuration can be associated to an $m$-element occupancy vector $\vec{s}(\vec{x})=(s_0,s_1,\ldots, s_{m-1})$, with each element $s_i\in{0,\ldots,n}$ indicating the number of photons in the $i$-th optical mode. If $m$ photons are injected into the $m$ distinct input modes of a $m\times m$ DFT, defined via the unitary $U_{j,k} = \exp(jk \frac{2\pi \I}{m})/\sqrt{m}$, the amplitude of a given $m$-photon configuration $\vec{x}$ at the output is given by the matrix permanent~\cite{Tichy2010}

\begin{align}
    \langle \Phi(\vec{x}) | \Psi \rangle &= \frac{1}{\sqrt{\prod_i s_i(\vec{x})!}} \sum_{\sigma \in P_{m}} \prod_{i =0}^{m-1} U_{x_i,\sigma_i} \nonumber \\
    &= \frac{1}{\sqrt{m^m \prod_i s_i(\vec{x})!}} \sum_{\sigma \in P_{m}} \exp\left[\left(\sum_{i =0}^{m-1}  \sigma_i x_i \right) \frac{2\pi\I}{m} \right].
\end{align}
In the main text we show that in the $N$-photon $d$-dimensional heralded GHZ state generator shown in Fig.~1 the ZTL implies that the output configurations with non-zero amplitudes are only $d$, and are of the following form: $m-N$ photons are the in the $0$-th mode, and the remaining photons are in a configuration $\vec{b}_k =( y_{k,1}, y_{k,2}, \ldots, y_{k,N})$, with $y_{k,i}\in \{1,\ldots,m-1\}$ and $\sum_{i=1}^m y_{k,i} = 0 \mod m$. Property 3 for the set $\chi$ also implies configurations satisfying $\sum_{i=1}^m y_{k,i} = m$.  Each of the $d$ possible configurations $\vec{b}_k$ is mapped directly into the logical state of the $N$ $d$-dimensional qudits: $\vec{b}_k \mapsto \ket{k,k,\ldots,k}$. In order to show that the states generated are in fact GHZ states, we will prove that the amplitude for each configuration $\vec{b}_k$ is the same. For a given $k\in\{0,\ldots,d-1\}$, the amplitude of the associated state is given by 
\begin{equation}
    \langle k,k,\ldots,k | \Psi \rangle = \frac{1}{\sqrt{m^m (m-N)!}} \sum_{\sigma \in P_{m}} \exp\left[\left(\sum_{i =0}^{N-1}  \sigma_i y_{k,i} \right) \frac{2\pi\I}{m} \right].
\end{equation}
Now, because $\sum_{i=1}^m y_{k,i} = m$ and  $y_{k,i}<m$ for all $k$ and $i$, we can estimate this sum over the permutations using Formula~\ref{Formula:GeneralP}, giving the amplitude
\begin{equation}
    \langle k,k,\ldots,k | \Psi \rangle = (-1)^{N-1} (N-1)! \sqrt{\frac{(m-N)!}{m^{m-2} }}.
\end{equation}
It is immediate to see that this quantity does not depend on $k$, which immediately implies that for any $N$ and $d$ the amplitudes are equal for all $k$, and therefore that the state is in fact an $N$-photon $d$-dimensional GHZ state. The success probability is given by 
\begin{equation}
\label{eq:succprob}
    P_{\text{succ}}(d,N) = \sum_{k=0}^{d-1} |\langle k,k,\ldots,k | \Psi \rangle|^2 = d (N-1)! ^2 \frac{[m(d,N)-N]!}{m(d,N)^{m(d,N)-2}}
\end{equation}
where we have now written explicitly the dependence of $m$ on both $d$ and $N$. Although, as discussed in Appendix~\ref{SupplInfo:GHZExamples}, for each $d$ and $N$ a value of $m$ that provides the desired GHZ structure exists, it is difficult to find an analytic dependence of $m$ from $d$ and $N$ in the general case. However, because the output circuit needs to encode $N$ photons in dimension $d$, necessarily $m>Nd$, meaning that $P_{\text{succ}}(d,N)$ in our scheme necessarily decreases exponentially with both $N$ and $d$. Such unfavourable scaling is due to the fact that, for simplicity in the calculations to show the generality, we are considering only one out of combinatorially many possible heralding combinations that can give raise to GHZ states. Approaches to explore all such combinations to exponentially boost the success probability are discussed in section~\ref{SupplInfo:ProbBoost}.

For the case of Bell states, i.e. $N=2$, a simple solution with $m=2d+1$ is shown in the main text, implying a success probability:
\begin{equation}
    P_{\text{succ}}(d,2) = d  \frac{(2d-1)!}{(2d+1)^{2d-1}}.
\end{equation}


\section{Examples of solutions for GHZ generation }

\label{SupplInfo:GHZExamples}

We here investigate solutions to the scheme to generate GHZ of $N$ photons in arbitrary dimension $d$ via DFT interferometers. Note first that, given a set $\mathcal{X}$, verifying whether it allows a partition $\mathcal{B}_\mathcal{X}$ that satisfies conditions 1-3 is a special case of the $k$-partitioning problem, and therefore is in general \texttt{NP}-hard~\cite{babel1998}. Nevertheless, as discussed below, general solutions for arbitrary $N$ and $d$ can be found. These, however, are likely to be highly not optimal in terms of auxiliary resources required and success probability. We also report more efficient solution for cases where $N$ and $d$ are small enough to make the optimisation task computationally feasible.  

\subsection{A general solution for GHZ generation}

We describe here in more detail the example provided in the main text of a general solution for $m$ and $\mathcal{X}$ that satisfies conditions 1-3 from the main text. The scope of this example is to show that our scheme based on linear optics and Fourier-transform interferometers provides a way to generate of $N$-photon $d$-dimensional GHZ states for any $N$ and $d$, but for simplicity we do not focus here on the efficiency of the protocol. In fact, solutions from this general example are very sub-optimal compared to the special case shown in the next section. We leave the improvement of success probabilities  and the reduction of resource requirements for general solutions as an interesting open question.

To find a solution for the generation of a GHZ state of $N$ photons and $d$ dimensions with our scheme, we need to find an integer $m$ and and set $\mathcal{X} \subseteq \{0,\ldots,m-1\}$ such that the set 
\begin{equation}
    \mathcal{B}_{\mathcal{X}}=\Big\{ \vec{b}_k = ( y_{k,1}, y_{k,2}, \ldots, y_{k,N} )\Big| \sum_{i=1}^N y_{k,i} = 0 \mod m,\  y_{k,i} \in \mathcal{X}
    \Big\}.
\end{equation}
satisfies conditions 1-3 from the main text. 

In order to build up a general solution, it is useful to note that, if we consider a solution $\mathcal{X}$ and remove all the terms $y_{k,N}$, then the sums $\sum_{i=1}^{N-1} y_{k,i} = m - y_{k,N}$ must all have different values. A simple, although possibly inefficient, way to ensure this is to start with a reduced set $\mathcal{Y} = \mathcal{X} \setminus \{y_{k,N}\}_{k}$ such that any combinations of $N-1$ elements from this set, including repetitions, provide different sums. The task of finding $\mathcal{Y}$ is then similar to the problem of finding sets of integers with distinct subset sums, which is a well-known problem in combinatorial number theory (note however that multisets need to be considered to include sums with repeated elements)~\cite{erdos1941,guy1982,bohman1998}. A simple solution for a set such that any combinations of $N-1$ elements from this set (including repetitions) provide different sums is given by $\mathcal{Y} = \{ N^0,N^1,\ldots,N^{(N-1)d - 1}\}$. In fact, different combinations correspond to different integer numbers in basis $N$, therefore providing different sums. We can then complete $\mathcal{Y}$ into $\mathcal{X}$ by setting $y_{k,i} = N^{id + k}$ for all $i\in [ 0, N-1]$ and $k \in [0, d-1]$, and defining the last $d$ elements as 
\begin{equation}
    y_{k,N} = m - S_k  = m - N^k\frac{ N^{(N-1)d} - 1}{N^d -1},
    \label{eq:analy_complete_set}
\end{equation}
where $S_k = \sum_{i=0}^{N-2} N^{id+k}$. This ensures all $\vec{b}_k$ sum up to $m$. We are now left to choose $m$. A challenge here is to pick $m$ so that no $N$-elements combinations of $\mathcal{X}$ sum up to multiples of $pm$ with $p>1$, and that no $y_{k,N}$ overlaps with the elements already in $\mathcal{Y}$. A way to do this is to pick $m> N \max_k S_k$, therefore we use

\begin{equation}
    m = N \max_k S_k + 1 = \frac{N^{Nd} - 1}{N^d - 1}.
\end{equation}
In conclusion, the resulting general solution is given by the set

\begin{equation}
    \mathcal{X} = \{ N^0,N^1,\ldots,N^{(N-1)d - 1}\} \cup \{m - N^k\frac{ N^{(N-1)d} - 1}{N^d -1}\}_{k \in [0, d-1]},
    \label{eq:analytic_sol}
\end{equation}
and $m= (N^{Nd} - 1)/(N^d - 1)$. 

The scope of this solution is to illustrate that for any $N$ and $d$ the generation of $N$-photon $d$-dimensional heralded GHZ states is possible with only linear optics using our scheme. However, we remark that this solution is far from optimal. 
Below we describe algorithms able to numerically find significantly better solutions for computationally practicable values of $N$ and $d$.
%

\subsection{Brute-force optimised solutions for small values of $N$ and $d$}
For small values of $N$ and $d$, optimal solutions for the heralded GHZ generation with our scheme can be found numerically using brute force approaches. The parameter to be optimised is the number of photons and modes $m$, which is also connected to the success probability via Eq.~\ref{eq:succprob}. The brute force search for solutions with minimal $m$ was performed as follows: we start from a value of $m=m_0$ large enough (we started from $m_0 = 40$ for the search of the solution in Fig.~1c); we then test the possible subsets $\mathcal{X}\subset \{0,\ldots,m\}$ with $Nd$, and check whether they satisfy conditions 1-3, i.e. if they represent valid solutions to the GHZ generation scheme. Because the number of such possible subsets increases combinatorially as ${m}\choose{Nd}$, if $m$ is too large for a reasonable computational time (in our case $m\gtrsim 30$), we used a limited number of randomly sampled subsets rather than all possible subsets. We empirically found that, for the values of $m$, $N$ and $d$ tested here, testing 10000 different random subsets was enough to find possible solutions with high probability, if solutions existed. If at least one solution is found for a given $m$, we decrease it by one ($m \mapsto m -1$) and repeat the process. If we don't find any solution for a given $m$, we conclude that the optimised value is $\bar{m}=m+1$, and a solution $\mathcal{X}$ is also given by one of those found at the step with $m=\bar{m}$. Using this brute force approach, in the case $(N,d)=(3,3)$ we found the optimised solution with $\bar{m}=25$ described in the main text and shown in Fig.~1c.

\subsection{
An algorithm for near-optimal solutions with larger $N$ and $d$
}

While the brute force method described above always provides optimised solutions for the proposed GHZ scheme (in terms of number of input photons $m$), it is very computationally costly and we found it not practicable to find solutions for larger values of $N$ and $d$. On the other hand, the general solutions from Eq.~\ref{eq:analytic_sol} readily provide analytical solutions for any $N$ and $d$, although highly sub-optimal and likely not suitable for experimental implementations. An approach which achieves a trade-off between these two methods, i.e. providing near-optimal solutions but with significantly better computational cost, is therefore worth investigating.  

We developed an algorithm to find solutions for the GHZ scheme which performs such a trade-off, and is readily able to find solutions for values $N\geq3$ and $d\geq3$ on a standard laptop, significantly improving the general analytical solutions of Eq.~\ref{eq:analytic_sol} in terms of resource optimisation and the brute-force approach in terms of computational cost. The pseudocode of the algorithm is provided in Algorithm~\ref{alg:DSI}, and a Python implementation is freely accessible on GitHub~\cite{lo_sim}. The algorithm, which we call Different-Sums Iterative (DSI) algorithm, proceeds in two parts. In the first part we iteratively build up a set $\mathcal{Y}$ of $(N-1)d$ distinct positive integers such that any sum of $N-1$ of its elements, possibly including repetitions, is different and such that $\max(\mathcal{Y})$ is minimal. This part is thus similar to the first part of the analytical approach, where $\mathcal{Y}$ was built analytically at the cost of large overheads in terms of $\max(\mathcal{Y})$. This optimised solution for $\mathcal{Y}$ is found iteratively: starting from the $(N-1)$-element set $\mathcal{Y}=\mathcal{Y}_0=\{1,\ldots,N-1\}$, at each step we find the smallest positive integer $\bar{a}$ such that, if it is included in the current set $\mathcal{Y}$, it preserves the desired condition (i.e. that any sum of $N-1$ of elements, including repetitions, is different). The number $\bar{a}$ is then added to $\mathcal{Y}$ and the process is repeated $(N-1)(d-1)$ times to get a set $\mathcal{Y}$ with $(N-1)d$ distinct elements. This part is fast enough to find sets $\mathcal{Y}$ with $N$ and $d \gtrsim 8$ on standard laptops.

\begin{table}
\flushleft
\begin{tabular}{|c|c||l|l|l|l|}
\hline
\multicolumn{2}{|c||}{\multirow{2.2}{*}{\shortstack[c]{GHZ\\ scheme}}} & \multicolumn{4}{c|}{Dimension $d$} \\ \cline{3-6}

\multicolumn{2}{|c||}{} & $d=2$ & $d=3$ & $d=4$ & $d=5$    \\ \hhline{|=|=#=|=|=|=|}
\multirow{17}{*}{\rotatebox[origin=c]{90}{Qudit number $N$}} & $N=3$  
& \begin{tabular}[c]{@{}l@{}} $m=15$ \vspace{4pt}\\  $\mathcal{X}=\{1, 2, 3,$\\ $4, 8, 12\}$\vspace{4pt}\\ $\mathcal{B}_\mathcal{X}=\{(1, 2, 12),$\\ $(3, 4, 8)\}$\end{tabular} 
&  \begin{tabular}[c]{@{}l@{}} $32$ \vspace{4pt}\\ $\{1, 2, 4, 7, 8, $\\ $11, 13, 21, 29\}$ \vspace{4pt}\\ $\{(8, 11, 13), (4, 7, 21), $\\$ (1, 2, 29)\}$\end{tabular} 
&  \begin{tabular}[c]{@{}l@{}} $58$ \vspace{4pt}\\ $\{1, 2, 4, 5, 6, 8, 13, $\\$21, 31, 41, 45, 55\}$\vspace{4pt}\\ $\{(1, 2, 55), (5, 8, 45), $\\$(4, 13, 41), (6, 21, 31)\}$\end{tabular} 
&  \begin{tabular}[c]{@{}l@{}} $109$ \vspace{4pt}\\ $\{1, 2, 4, 7, 8, 13, 21, 30, 31, $\\$33, 45, 66, 81, 97, 106\}$\vspace{4pt}\\ $\{(4, 8, 97), (31, 33, 45), (1, 2, 106), $\\$ (7, 21, 81), (13, 30, 66)\}$\end{tabular}
 \\ \cline{2-6} 
& $N=4$   
& \begin{tabular}[c]{@{}l@{}} $m=69$ \vspace{4pt}\\  $\mathcal{X}=\{1, 2, 3, $\\$4, 8, 23, 43, 54\}$\vspace{4pt}\\ $\mathcal{B}_\mathcal{X}=\{(1, 2, 23, 43), $\\$(3, 4, 8, 54)\}$\end{tabular}  
& \begin{tabular}[c]{@{}l@{}} $292$ \vspace{4pt}\\  $\{1, 2, 3, 4, 8, 23,  51, $\\$ 54, 88, 164, 201, 277\}$\vspace{4pt}\\ $\{(23, 51, 54, 164), $\\$(3, 4, 8, 277), $\\$ (1, 2, 88, 201)\}$\end{tabular}  
& \begin{tabular}[c]{@{}l@{}} $1030$ \vspace{4pt}\\  $\{1, 2, 3, 4, 8, 23, $\\$54, 88, 133, 164, 277, $\\$279, 456, 609, 1004, 1015\}$\vspace{4pt}\\ $\{(3, 4, 8, 1015), $\\$(133, 164, 277, 456), $\\$(54, 88, 279, 609), $\\$(1, 2, 23, 1004)\}$\end{tabular} 
& -
\\ \cline{2-6} 
& $N=5$   
& \begin{tabular}[c]{@{}l@{}} $m=391$ \vspace{4pt}\\  $\mathcal{X}=\{1, 2, 3, 4, 14, 20, $\\$54, 138, 196, 350\}$\vspace{4pt}\\ $\mathcal{B}_\mathcal{X}=\{(1, 2, 54, 138, 196), $\\$(3, 4, 14, 20, 350)\}$\end{tabular}  
& \begin{tabular}[c]{@{}l@{}} $5920$ \vspace{4pt}\\  $\{1, 2, 3, 4, 14, 20, 54, $\\$138, 209, 350, 815, 1781, $\\$2977, 5509, 5883\}$\vspace{4pt}\\ $\{(1, 2, 14, 20, 5883), $\\$(3, 4, 54, 350, 5509), $\\$(138, 209, 815, 1781, 2977)\}$\end{tabular}
& -
& -
\\ \hline
\end{tabular}

\vspace{10pt}

\begin{tabular}{|c|c||l|l|l|}
\hline
\multicolumn{2}{|c||}{\multirow{2.2}{*}{\shortstack[c]{GHZ\\ scheme}}} & \multicolumn{3}{c|}{Dimension $d$} \\ \cline{3-5}

\multicolumn{2}{|c||}{} & $d=6$ & $d=7$ & $d=8$  \\ \hhline{|=|=#=|=|=|}

\rotatebox[origin=c]{90}{Qudit number $N$} 
& $N=3$  
& \begin{tabular}[c]{@{}l@{}} $m=154$ \vspace{4pt}\\  $\mathcal{X}=\{[1, 2, 4, 7, 8, 13,  18, $\\$ 21, 26, 31, 45,  66, 81,  88, $\\$ 97, 123, 142, 151]\}$\vspace{4pt}\\ $\mathcal{B}_\mathcal{X}=\{(7, 66, 81), (4, 8, 142), $\\$(26, 31, 97), (21, 45, 88), $\\$(1, 2, 151), (13, 18, 123)\}$\end{tabular} 
&  \begin{tabular}[c]{@{}l@{}} $219$ \vspace{4pt}\\ $\{1, 2, 4, 5, 8, 13, 15, 16, 21, $\\$31, 45, 66, 77, 81, 97, 123, $\\$148, 175, 182, 207, 216\}$ \vspace{4pt}\\ $\{(5, 66, 148), (13, 31, 175), (45, 77, 97), $\\$(4, 8, 207), (16, 21, 182), $\\$ (1, 2, 216), (15, 81, 123)\}$\end{tabular} 
&  \begin{tabular}[c]{@{}l@{}} $263$ \vspace{4pt}\\ $\{1, 2, 3, 4, 8, 13, 21, 28, 31, 34, $\\$36, 43, 45, 66, 81, 97, 123, 148, $\\$176, 182, 204, 246, 252, 260\}$\vspace{4pt}\\ $\{(34, 81, 148), (43, 97, 123), (21, 66, 176), $\\$(3, 8, 252), (28, 31, 204), (1, 2, 260), $\\$(4, 13, 246), (36, 45, 182)\}$\end{tabular} 
\\ \hline 
\end{tabular}
\caption{
Exemplary list of solutions obtained via the DSI algorithm to the GHZ scheme for different values of $(N,d)$. For each case we report, from top to bottom, the number of modes and input photons $m$, the list of encoding modes $\mathcal{X}$, and the associated set $\mathcal{B}_\mathcal{X}$.
}
\label{tab:DSI}
\end{table}

\begin{figure}
    \centering
    \includegraphics[width=0.5\linewidth]{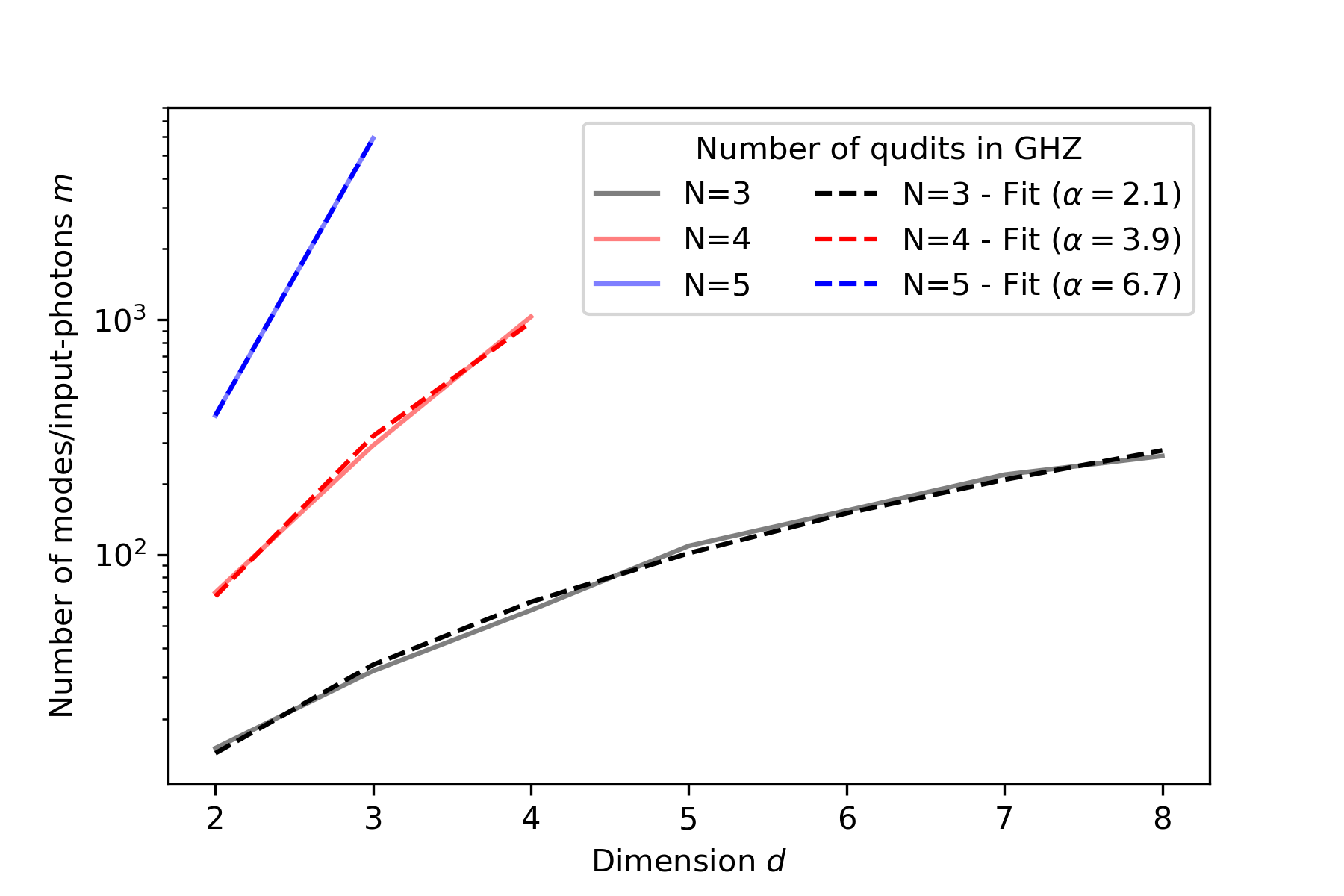}
    \caption{
    Scaling of the number of input photons $m$ in the GHZ generation scheme given by the solutions of the DSI algorithm. Solid lines show calculated values of $m$ for different qudit numbers in the GHZ of dimensionality $d$: $N=3$ (black), $N=4$ (red) and $N=5$ (blue). Dashed lines are polynomial fits $m(d)_N = \gamma_N + \beta_N d^{\alpha_N}$, with the exponents fit parameters $\alpha_N$ reported in the legend for the different $N$ values. 
    }
    \label{fig:GHzscheme_NewAlg}
\end{figure}

To complete the set $\mathcal{Y}$ to a possible solution of $Nd$ qudit-encoding modes for the GHZ scheme, we can proceed similarly to the second part of the analytical approach (Eq.~\ref{eq:analy_complete_set}): because each partition $P=\{P_1,\ldots,P_d\}$ of $\mathcal{Y}$ into $d$ sets of $N-1$ elements ($|P_i|=N-1$) will contain subsets $P_i$ with different sums $\sigma_i = \sum_{p\in P_i}p \neq \sigma_j$, choosing an additional element $\bar{y}_i = m-\sigma_i$ we get that each subset $P_i \cup \{\bar{y}_i\}$ will now sum up to $m$. We thus obtain the set $\mathcal{X} = \mathcal{Y} \cup \{\bar{y}_1,\ldots,\bar{y}_d\}$ as a possible solution. The challenge is to find the smallest $m$ such that  $\mathcal{B}_\mathcal{X}$ for the obtained set $\mathcal{X}$ satisfies conditions 1-2 of the GHZ scheme, i.e. such that $m$ and $\mathcal{X}$ form a valid solution. For example, some $\bar{y}_i$ could already be in $\mathcal{Y}$, in which case  the obtained $\mathcal{B}_\mathcal{X}$ would not satisfy conditions 1 and 2 for a valid solution to the GHZ scheme. With this approach, not all sets obtained $\mathcal{X}$ will thus necessarily be correct solutions, and we therefore need a verification step which is repeated until a valid solution is found. In the pseudocode provided in Algorithm~\ref{alg:DSI}, this part is done via brute-force starting from smaller values of $m$ and looking if any of the possible valid partitions of $\mathcal{Y}$ would give raise to a set $\mathcal{B}_\mathcal{X}$ which satisfies conditions (1-3) of the GHZ scheme. This is repeated, increasing the value of $m$ until a correct solution $\mathcal{X}$ is found, and the obtained values of $\mathcal{X}$ and $m$ are finally returned.

 Using the DSI algorithm we can easily obtain solutions for high-dimensional heralded GHZ state generation with up to $N=5$ photons and, for $N=3$, up to dimensionality $d=8$. Examples for sets $\mathcal{X}$ and the number of modes $m$ for such solutions are reported in Supplementary Table~\ref{tab:DSI}. For $N=2$ the algorithm provides the same optimised solutions for any $d$ as the ones reported in the main text (see Fig.1b). For $(N,d)=(3,3)$ the solution found requires $m=32$, which is slightly worse than the optimised solution with $m=25$ obtained with the brute-force approach (see Fig.1c). This indicates that, while the solutions given by the DSI algorithm are likely close to the optimised ones, they are in general non-optimal in terms of $m$. To further investigate this, Supplementary Fig.~\ref{fig:GHzscheme_NewAlg} shows in log scale the values of $m$ for the solutions given by the DSI algorithm for different values of $N$ and $d$. The curve for $N=3$ shows a sub-exponential increase of $m$ with the dimension (in particular, a scaling $m \sim d^2$ for $N=3$), which significantly improves the exponential increment of the analytical solution in Eq.~\ref{eq:analytic_sol}, potentially making these schemes significantly more suitable for practical implementations.

We finally comment on possible routes to improve the DSI algorithm. First, we note that in the first part of the DSI algorithm in obtaining the set $\mathcal{Y}$ we insist that any different choice of $N-1$ of its elements, including repetitions, must have a different sum. Such a constraint can be relaxed. In fact, a looser condition for the set $\mathcal{Y}$ is that there exists a partition  $P=\{P_1,\ldots,P_d\}$ into $d$ sets of $N-1$ elements such that each of the sums $\sigma_i = \sum_{p\in P_i}p$ cannot be obtained by any other choice of $N-1$ of elements of $\mathcal{Y}$ (including repetitions) that is not $P_i$. Such a condition is sufficient to complete $\mathcal{Y}$ to a possible solution $\mathcal{X}$ as in the current algorithm. We believe this relaxation in obtaining $\mathcal{Y}$ is likely to provide better solutions in terms of the number of required resources $m$. In terms of computational resources required to find a solution, we note that the brute-force method to perform the second part of the DSI algorithm is the one that requires more memory usage and computational time, in particular when exploring all the possible partitions of $\mathcal{Y}$. Finding heuristics that require to explore fewer partitions of $\mathcal{Y}$ are likely to improve this part significantly and enable the investigation of solutions for larger values of $N$ and $d$.

\begin{algorithm}[H]
\caption{
Different-Sums Iterative (DSI) algorithm
}\label{alg:DSI}
\begin{algorithmic}[1]
\Require Number of qudits in the GHZ $N$, dimensionality for each qudit $d$.
\State $M \gets N-1$
\State $\mathcal{Y} \gets \{1,\ldots,M\}$ \Comment{Initialize the set $\mathcal{Y}$.}
\For{$k\in 1 \rightarrow M$}
\State $S_k\gets\{y_1 + ... + y_k | y_i \in \mathcal{Y}\}$ \Comment{Calculate set of sums $S_k$ of all choices of $k$ elements from $\mathcal{Y}$ (with repetitions).}
\EndFor

\For{$n\in M \rightarrow Md -1$} \Comment{Iteratively adds elements to the set $\mathcal{Y}$ such that any subset of $M$ elements, including repetitions, always has different sum.}
\State $\Gamma \gets \{s_M - s_{M-1} | s_n \in S_M , s_{M-1} \in S_{M-1}, s_M - s_{M-1}>0 \}$
\State $\bar{a} \gets \min(\mathbb{N} \setminus (\Gamma \cup \mathcal{Y})$ \Comment{Find new element $\bar{a}$ to add to $\mathcal{Y}$ as the minimum positive integer not in $\Gamma$ nor already in $\mathcal{Y}$.}
\State Append $\bar{a}$ to $\mathcal{Y}$ \Comment{Update $\mathcal{Y}$ including new element $\bar{a}$.}
\State Append $\bar{a}$ to $S_1$
\For{$k\in 2 \rightarrow M$}
\State $S_k \gets S_k \cup \{\sigma_{k-1} + \bar{a} | \sigma_{k-1} \in S_{k-1}\}$ \Comment{Iteratively update sets of sums $S_k$ after $\bar{a}$ is added to $\mathcal{Y}$.}
\EndFor
\EndFor 
\State $\mathcal{P} \gets \{ (P_1,\ldots,P_d) | (P_1,\ldots,P_d) \text{ is a partition of } \mathcal{Y} \text{ in } d \text{ subsets with } |P_i|=M \text{ elements.}\}$
\State $\Sigma \gets \{ \left(\sum_{p \in P_1} p,\ldots,\sum_{p \in P_d} p\right) | (P_1,\ldots,P_d) \in \mathcal{P} \}$ \Comment{Given how $\mathcal{Y}$ is constructed, each sum $\sigma_i=\sum_{p \in P_i} p$ is different.}
\For {$m\in \max(\mathcal{Y})+1 \rightarrow m_{\text{max}}$} \For {$\sigma \in \Sigma$}
\State $\mathcal{X} \gets \mathcal{Y} \cup \{ m - \sigma_i | \sigma_i \in \sigma \}$ \Comment{Complete the set with $d$ different elements such that the partitions sum to $m$.}
\State Calculate $\mathcal{B}_\mathcal{X}$ given $\mathcal{X}$ and $m$.
\If{$\mathcal{B}_\mathcal{X}$ satisfies conditions (1-3) of the GHZ scheme}
\State \textbf{return} $\mathcal{X}$ and $m$ \Comment{If the obtained $\mathcal{B}_\mathcal{X}$ satisfies the GHZ scheme conditions, return $\mathcal{X}$ and $m$ as a solution.}
\EndIf
\EndFor
\EndFor
\end{algorithmic}
\end{algorithm}

\subsection{Explicit explanatory construction of the GHZ scheme}

We describe here an explanatory example of how solutions for the GHZ generation scheme are mapped into the qudit encodings. In particular, we consider the $(N=3, d=3)$ GHZ state. As reported above, we have an optimized solution of the associated combinatorial number theory problem given by $m=25$ and $\mathcal{X}=\{1,2,3,4,5,9,13,16,22\}$, also represented in Fig.~1 of the main text. We construct the set $\mathcal{B}_\mathcal{X}$ given by all possible choices of $N$ elements from $\mathcal{X}$, including possible repetitions, such that their sum is a multiple of $m$. The set $\mathcal{B}_\mathcal{X}$ thus represents the set of output photon configurations allowed by the ZTL. For the values of $m$ and $\mathcal{X}$ as above, one can see that the only choices of $N$ elements summing up to multiples of 25 are $\mathcal{B}_\mathcal{X}=\{(1,2,22), (3,9,13), (4,5,16) \}$. This set $\mathcal{B}_\mathcal{X}$ trivially satisfies conditions 1-3 for the GHZ generation specified in the main text, indicating that $m$ and $\mathcal{X}$ in fact provide a valid solution. 

Proceeding as described in the main text, we associate to each integer $\ell\in\mathcal{X}$ the $\ell$-th output mode of the DFT interferometer. From the set $\mathcal{B}_\mathcal{X}=\{(1,2,22), (3,9,13), (4,5,16) \}=\{(y_{1,1}, y_{1,2}, y_{1,3}), (y_{2,1}, y_{2,2}, y_{2,3}),(y_{3,1}, y_{3,2}, y_{3,3}) \}$ we obtain the logical encoding of the qudits into the optical modes as $\ket{k}_i \leftrightarrow y_{k,i}$, where $\ket{k}_i$ labels the logical state $k\in\{0,\ldots,d-1\}$ of the $i$-th qudit. The $N=3$ qudits are thus defined in the modes $Q_1=\{1,3,4\}$, $Q_2=\{2,9,5\}$, $Q_3=\{22,13,16\}$. The set of $m-Nd=16$ heralding modes is the complementary set of $\mathcal{X}$, i.e. $\bar{\mathcal{X}}=\{0,6,7,8,10,11,12,14,15,17,18,19,20,21,23,24\}$. The modes in $\bar{\mathcal{X}}$ are thus detected to check if the $m-N=22$ heralding photon are in a valid heralding configurations (e.g. all 22 photons in the 0-th mode and vacuum in all other modes in $\bar{\mathcal{X}}$). If a valid heralding configuration is detected, the ZTL ensures that the three output photons can only be one of the configurations in $\mathcal{B}_\mathcal{X}$. Furthermore, condition 3 implies that we actually have a uniform superposition between these configurations (see Formula~3 in Appendix 1). Using the mapping as above we thus finally obtain that the output state corresponds to $\ket{\text{GHZ}_{(3,3)}}=(\ket{000}+\ket{111}+\ket{222})/\sqrt{3}$, as desired.

\section{Boosting the success probability\label{SupplInfo:ProbBoost}}

We here describe the methods used to boost the success probability of the heralded high-dimensional GHZ state generation. The main idea is, for a given solution $\mathcal{X}$, to include additional heralding events in the heralding modes $\bar{\mathcal{X}}$ which, although difficult to treat in the general case, can still provide GHZ entanglement for a specific solution $\mathcal{X}$. In fact, while in the main text we mainly focus on the single heralding event with all the heralding photons in the zero-th mode, which simplified proving the generality of our scheme, many more heralding event are likely to be valid for a specific $\mathcal{X}$. 

Specifically, if a heralding pattern $(e_1,\ldots,e_{m-N})$ is observed ($e_i \in \bar{\mathcal{X}}$) such that 
\begin{equation}
    \sum_{i=1}^{m-N} e_i = 0\mod m,
    \label{eq:AlmostGHZ}
\end{equation}
then, following the same reasoning as in the main text, conditions 1 and 2 ensure (via the ZTL) that the output state is a superposition of the logical states $\ket{k,\ldots,k}$, $k \in 0,\ldots,d-1$. 
The additional heralding patterns now considered are thus all configuration of modes that sum up to a multiple of the number of modes.
However, 
because condition 3 is no longer necessarily satisfied,
the amplitudes in the superposition might not be uniform. On the other hand, such states can be probabilistically corrected to GHZ qudit states using heralded Procrustean distillation, similar to the schemes shown by Joo et al.~\cite{Joo2007}. Amplitudes are balanced by passing the optical modes of one of the qudits through $d-1$ beam splitters with appropriately chosen reflectivities and heralding vacuum on one of the output ports of the each of the beam splitters. This correction process is heralded and its success probability corresponds to $d \min_k |a_k|^2$, where $a_k$ is the amplitude of the term $\ket{k,\ldots,k}$ in the superposition. Given an observed heralding pattern at the output of the DFT satisfying Eq.~\ref{eq:AlmostGHZ}, reconfiguring a Procrustean distillation circuit acting on only one of the qudits allows heralding of successful GHZ state generation.  

While this approach thus requires additional resources and feed-forward, combinatorially many heralding patterns can now be used, instead of a single one, to enormously boost the success probability for generating GHZ qudit states. For example, for the $(N,d)=(3,3)$ scheme shown in Fig.~1c, with the fixed heralding set of modes $\bar{\mathcal{X}}= \{0,6,7,8,10,11,12,14,15,17,18,19,20,21,23,24\}$ there are $>10^7$ possible heralding patterns of the $23$ heralding photons that satisfy Eq.~\ref{eq:AlmostGHZ}. 

Estimating the exact total success probability for the scheme taking into account Procrustean distillation would require calculating the output amplitudes for all such heralding events at the output of the DFT. Given that this requires simulating the evolution of 25 photons for each configuration, it is not computationally viable. Instead, we use a Monte Carlo approach to estimate it. Specifically, we sampled $32\times 10^5$ output configurations from the DFT using the algorithm of Clifford and Clifford~\cite{Clifford2018}, and stored the outputs where the heralding pattern on the modes $\bar{\mathcal{X}}$ satisfied condition Eq.~\ref{eq:AlmostGHZ}. For such successful heralding pattern, we calculated the amplitudes $a_k$ associated to the terms $\ket{k,\ldots,k}$ in the qudits superposition. Each pattern was then accepted with probability $d \min_k |a_k|^2$, corresponding to the success probability of the Procrustean distillation. The estimation of the total success probability was finally obtained as the ratio between the number of accepted configuration and the total number of samples tested. Repeating the procedure also provides an estimation for the uncertainty of the estimate. For the $(N,d)=(3,3)$ scheme in Fig.~1c we obtained an estimate of the success probability of $0.8(1)\times 10^{-4}$, a six orders of magnitude improvement with respect to the case with a single heralding pattern. 

To understand the practicability on near-term hardware for the schemes after this further success probability optimisation, we investigate the expected rates for the $(N,d)=(3,3)$ GHZ state generator with state-of-the-art hardware. In recent work from quantum dot research groups, a single photon generation efficiency over 92\% has been reported with current technology~\cite{Uppueabc8268}. The state of the art for low-loss multiport interferometers is 99\% transmission~\cite{Zhong1460} and state of the art number resolving detectors have 98\% efficiency~\cite{Fukuda:11}. If we optimistically assume that these numbers can be achieved in a single experiment, which runs at the repetition rate of the quantum dot (145 MHz), we find that the optimised $0.8\times10^{-4}$ success probability corresponds to 677 successfully heralded and detected GHZ states per second.

The rules for generating GHZ states outlined in the main text allow us to prove that linear optical GHZ generation is possible for any $(N,d)$. However, we have shown that as we relax these requirements, huge performance improvements can be found. We therefore expect that by further relaxing these requirements (i.e. different unitary matrices, different number of input photons), further improvements can be found for all $(N,d)$.

\section{Generating other entanglement structures with the DFT-based scheme}

In the main text we focused on using the Zero-Transmission Law in DFT interferometers to engineer the GHZ states generation. We here show that the DFT-based scheme can also be used to directly generate a large variety of $N$-photon states in any dimensionality. In particular, we show that for a large variety of $N$-photon qudit states, using the DFT-based scheme the task of generating them can be reduced to a combinatorial number theory problem similar to the one for GHZ state generation.

Let's for example consider $N$-photon $d$-dimensional states $\ket{\psi}$ which, in the computational basis, can be written as a uniform $M$-terms expansion
\begin{equation}
    \ket{\psi}=\sum_{k=1}^M \ket{x_{k,1}, \ldots,x_{k,N}} / \sqrt{M}.
    \label{eq:UniState}
\end{equation}
Proceeding as for the GHZ generation scheme, one can see that if we inject $m$ input photons into an $m$-mode DFT and we can find a set $\mathcal{X}$ of $Nd$ output modes such that 
\begin{equation}
    \mathcal{B}_\mathcal{X} = \{ (y_{k,1}, \ldots,y_{k,N})\}_{k=1}^M
\end{equation}
and the mapping between the optical modes and the logical state is $x_{k,i} \leftrightarrow y_{k,i}$, the ZTL ensures that the state $\ket{\psi}$ is generated at the output. The task of generating $\ket{\psi}$ is therefore similar to the GHZ-scheme one, where now conditions 1-2 are generalised to require that $\mathcal{B}_\mathcal{X}$ has the same structure as the computational basis expansion of the state in Eq.~\ref{eq:UniState}. Condition 3 is still useful to ensure that the coefficients are real positive and uniform. 

Because the problem of generating states as in Eq.~\ref{eq:UniState} is very similar to the GHZ-scheme one, we can then use the same techniques describe in Appendix~3 to find valid solutions for $m$ and $\mathcal{X}$. For example, we can proceed by finding a set of $(N-1)d$ positive integers such that if we choose $N-1$ of them (including possible repetitions), we obtain different sums for different choices, as in the analytical solution (Eq.~\ref{eq:analytic_sol}) or in the DSI algorithm (Algorithm~\ref{alg:DSI}). 
We can then assign a mapping $x_{k,i} \leftrightarrow y_{k,i}$  (with $i\in\{1,\ldots,N-1\}$) between the logical states and the integers in the set, which represent the optical modes, and build up the set $\{(y_{k,1}, \ldots,y_{k,N-1})\}$ with the same structure of the target state $\ket{\psi}$ excluding the terms $x_{k,N}$. One can finally complement this set with the final $d$ modes taking $y_{k,N} = m - \sum_{i=1}^{N-1} y_{k,i}$ choosing $m$ as in the analytical solution (Eq.~\ref{eq:analytic_sol}) or in the DSI algorithm (Algorithm~\ref{alg:DSI}).

States of the form as in Eq.~\ref{eq:UniState} include states with complex asymmetric entanglement structures that arise for high-dimensional systems. For tripartite systems, such structures have been characterised, for example, via Schmidt Rank Vectors~\cite{Huber2013}. The heralded generation of such states directly via the DFT-based scheme could thus also enables the use of more complex entanglement resources for high-dimensional quantum information processing.

For example, to generate the asymmetric state $\ket{\psi_{332}}=(\ket{000}+\ket{111}+\ket{122})/\sqrt3$ from Ref.~\cite{Huber2013}, we used the brute-force numerical approach as in Appendix 3 to find the following optimised solution for the DFT-based scheme: $m=22$, $\mathcal{X}=(1, 2, 5, 6, 7, 9, 13, 16, 17)$, with the mapping between the logical states and the modes done via the qudit encodings $Q_1=(1, 7, 17)$, $Q_2=(5, 2, 6)$, $Q_3=(16, 13, 9)$. In fact, the only combinations of elements from $\mathcal{X}$ that sum up to multiples of $m$ (i.e. output combinations allowed by the ZTL) are $\mathcal{B}_\mathcal{X}=\{(1, 5, 16), (7, 2, 13), (7, 6, 9)\}$, which has the same structure as the $\ket{\psi_{332}}$ state in the computational basis. The ZTL thus generates at the output a uniform superposition of having the three photons in modes $(1, 5, 16)$, $(7, 2, 13)$, or $(7, 6, 9)$, which, applying the qudit encoding, corresponds the target logical state $\ket{\psi_{332}}$. Note also that to generate $\ket{\psi_{332}}$ state fewer resources are required with respect to the $(3,3)$ GHZ state, namely $m=22$ instead of $m=25$, which we associate to less entanglement being generated.

Furthermore, note that any state which, in the computational basis, has a $M$-terms expansion with real and positive amplitudes
\begin{equation}
    \ket{\psi}=\sum_{k=1}^M c_k \ket{x_{k,1}, \ldots,x_{k,N}},\qquad c_k\in \mathbb{R}^+
    \label{eq:DFTgeneratable}
\end{equation}
can always be generated by preparing the uniform superposition of the $M$ terms in Eq.~\ref{eq:UniState}
and than using Procrustean distillation to tailor the amplitude of each term to $c_k$~\cite{Joo2007}. Therefore, including local operations and detection, the high-dimensional entanglement structures possible to generate via the DFT-based scheme includes all states of the form in Eq.~\ref{eq:DFTgeneratable} and their local-equivalence classes. We believe this is likely to be further generalisable to any $N$-photon $d$-dimensional state, meaning that DFT-based schemes might also be universal, but leave this as an open question. On the other hand, note that the number of resources required to generates such states with the DFT-based scheme scales in the same way as for the GHZ case (see Eq.~\ref{eq:succprob}), meaning that exponentially many resources are required when increasing $N$. For a scalable architecture, the LOQC approach based on multiplexing the generation of small building blocks with small $N$ and fusing them together, as described in the main text, remains the only current option.

\section{Heralded vs. post-selected high-dimensional entanglement generation schemes}
As discussed in the introduction of the main text, most experimental and theoretical work on qudit entanglement has relied on post-selection. Post-selected entanglement is generated when the photons are destructively measured. This means that once the entanglement has been generated, it cannot be used in further quantum information processing steps. It is therefore not a viable resource for scalable quantum computation. While post-selected entanglement can still be a useful tool in other quantum technologies, such as in quantum communication and tests of non-locality, these areas are not the focus of this work.

In contrast, heralded entanglement does not destroy the photons in the entangled state and so provides ‘event-ready’ entanglement. This allows for successfully entangled photons to undergo subsequent evolutions. One important tool this enables is multiplexing of heralded processes. Here, many heralded processes are attempted in parallel and states which are successfully generated are routed to the desired modes using optical switching and feedforward. Multiplexing of heralded processes is a key tool in enabling scalable linear optical quantum computing, and is possible only in the absence of post-selection.

Furthermore, there has been interesting work which suggests that post-selected entanglement generation is capable of generating only a very limited portion of quantum states. See for example Ref.~\cite{Adcock_2018} for a study on which graph states cannot be postselected and Ref.~\cite{Krenn_2017} where they suggest that postselection may not be able to generate exact GHZ states for $d=3$, $N>4$ due to so-called `maverick terms'. On the other hand, since heralding must measure partial information about the state to detect a successful entanglement generation, in general heralded schemes require additional overheads with respect to post-selected ones (if post-selected schemes exist).

In Table~\ref{tab:post} we show the resources used in GHZ generation with post-selected schemes and our heralded scheme. While our scheme enables the heralded generation of any state, blank entries show cases where there is no known solution with post-selection. The overheads for the heralded scheme can be seen to become more significant when increasing $N$. On the other hand, note that for the constructions used in this work we mostly use $N\leq3$, where the overheads are still relatively small.

\begin{table}
\begin{tabular}{|l||l|l|l||l|l|l|}
\hline
\multirow{2}{*}{} & \multicolumn{3}{c||}{Post-Selected} & \multicolumn{3}{c|}{Heralded} \\ \cline{2-7} & $d=2$ & $d=3$  & $d=4$ & $d=2$ & $d=3$ & $d=4$ \\ \hhline{|=#=|=|=#=|=|=|}
$N=2$               & \begin{tabular}[c]{@{}l@{}}4 modes\\ 2 photons\end{tabular} & \begin{tabular}[c]{@{}l@{}}6 modes\\ 2 photons\end{tabular}  & \begin{tabular}[c]{@{}l@{}}8 modes\\ 2 photons\end{tabular} & \begin{tabular}[c]{@{}l@{}}5 modes\\ 5 photons\end{tabular}   & \begin{tabular}[c]{@{}l@{}}7 modes\\ 7 photons\end{tabular}     & \begin{tabular}[c]{@{}l@{}}9 modes\\ 9 photons\end{tabular}     \\ \hline
$N=3$               & \begin{tabular}[c]{@{}l@{}}6 modes\\ 3 photons\end{tabular} & \begin{tabular}[c]{@{}l@{}}9 modes\\ 3 photons\end{tabular}  & -                                                           & \begin{tabular}[c]{@{}l@{}}15 modes\\ 15 photon\end{tabular}  & \begin{tabular}[c]{@{}l@{}}25 modes\\ 25 photons\end{tabular}   & \begin{tabular}[c]{@{}l@{}}58 modes\\ 58 photons\end{tabular}     \\ \hline
$N=4$               & \begin{tabular}[c]{@{}l@{}}8 modes\\ 4 photons\end{tabular} & \begin{tabular}[c]{@{}l@{}}12 modes\\ 4 photons\end{tabular} & -                                                           & \begin{tabular}[c]{@{}l@{}}69 modes\\ 69 photons\end{tabular} & \begin{tabular}[c]{@{}l@{}}292 modes\\ 292 photons\end{tabular} & \begin{tabular}[c]{@{}l@{}}1030 modes\\ 1030 photons\end{tabular} \\ \hline
\end{tabular}
\caption{
Comparison of resources for post-selected and heralded entanglement generation schemes
}
\label{tab:post}
\end{table}

\section{Alternative qudit Bell pair generation circuits \label{bell_states}}

We here describe additional linear optical circuits for the generation of qudit Bell pairs which we identified while developing the general circuits described in the main text. These circuits can provide some improvements over the circuit shown in Fig.~1b of the main text. However, as far as we know, they do not generalise to the generation of GHZ-like entanglement for arbitrary $(N,d)$. All results in this section were found using our recently developed linear optical circuit simulator `lo\_sim'~\cite{lo_sim}. Example simulations of these circuits can be found in the lo\_sim repository.

\begin{figure}
    \centering
    \includegraphics[width=0.5\linewidth]{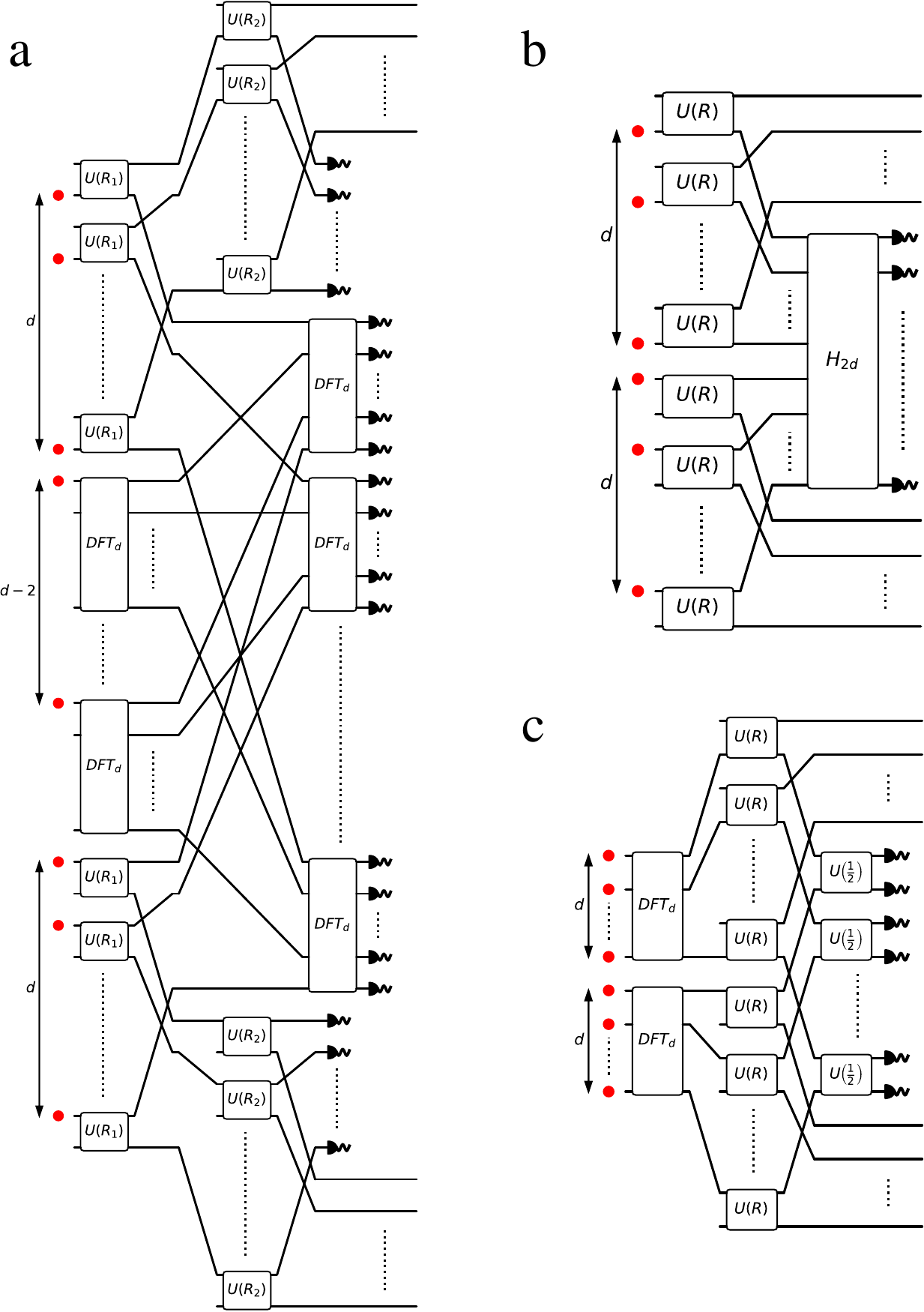}
    \caption{Schematic of alternative circuits for heralded qudit Bell pair generation. To maximise success probability for version 1 (a), we set $R_1 = 1/d$ and $R_2 = (d-2) /(d-1)$. For version 2 (b), we set $R=(d-1)/d$. For version 3 (c), we set $R=(d-1)/d$.}
    \label{bell_state_circuits}
\end{figure}

Throughout this section we describe a beam splitter with reflectivity $R$ by the matrix:

\begin{equation}
    U(R) = \begin{pmatrix}
    \sqrt{R} & i \sqrt{1-R} \\[6pt]
    i \sqrt{1-R} & \sqrt{R}
    \end{pmatrix}.
\end{equation}
To assess these circuits, it is here useful to adopt three different definitions of success probability, $p_a$, $p_f$, $p_c$, based on the hardware requirements of the generation scheme. To understand these different definitions, consider the qubit Bell generation circuit by Zhang et al.~\cite{Zhang2008}, which can produce Bell pairs which are defined over different allocations of modes to the two qubits. However, considering all of these outcomes as successes would be inappropriate for some applications. For this reason, we define success probabilities for cases with fixed allocations, $p_f$, and arbitrary allocations, $p_a$. Fixed allocation success probability describes the probability of heralding a Bell pair when we fix which modes are allocated to each of the two qudits. Arbitrary allocation success probabilities defines the probability of heralding a Bell pair where the qudits can be described by any partition of modes. For arbitrary allocated success, it is likely that an optical switch would be required to rearrange the modes based on some heralding outcome. 

As in Appendix~\ref{SupplInfo:ProbBoost}, in some cases, the circuits described in this section can produce states which have entanglement with full Schmidt rank, but which are not maximally entangled due to unbalanced amplitudes. Such states can be probabilistically corrected to maximally entangled qudit Bell pairs using Procrustean distillation. To include this possibility, we also define a `corrected' success probability, $p_c$. Corrected success describes the total success probability of the gate if such corrections are allowed. These correction operations include the deterministic correction of the mode allocation of the qudits, and therefore we always have $p_c \ge p_a \ge p_f$.

\subsection{Heralded Bell circuit: Version 1}
The heralded Bell pair generation circuit of Zhang et al.~\cite{Zhang2008} uses a type-II fusion gate at its core. We aimed to generalise this circuit for creating heralded qudit Bell states by using the type-II fusion gate from Luo et al.~\cite{Luo2019}. The layout of this circuit is shown for arbitrary dimension in Fig.~\ref{bell_state_circuits}a. Successful generation events are heralded by one photon being detected at the output of each DFT and a further $d-2$ photons being detected in both the first and last $d$ detectors.

The other circuits we present in the main text require $\le 2d+1$ photons, whereas this circuit requires $3d - 2$ single photon inputs and has success probabilities lower than other circuits we present, as shown in Table~\ref{probs_table}. We therefore believe this circuit to be of less practical interest compared to the other circuits we present.

\subsection{Heralded Bell circuit: Version 2 \label{v2}}
This circuit comes from a more direct generalisation of the Bell pair circuit of Zhang et al.~\cite{Zhang2008}. This circuit uses $2d$ photons -- which we conjecture to be the minimum number of photons required to produce a $d$-dimensional Bell state. The unitary, $H_{2d}$, shown in Fig.~\ref{bell_state_circuits}b, can be chosen to be any $2d \times 2d$ complex Hadamard. In Table~\ref{probs_table} we assess this circuit for the following three cases (labelled A, B, C):

\begin{equation}
    H_{2d, A} = DFT_{2d}
    \label{A}
\end{equation}

\begin{equation}
    H_{2d, B} =  \left(DFT_d\right)^{\oplus 2} \cdot P \cdot \left(DFT_2\right) ^ {\oplus d}
    \label{B}
\end{equation}

\begin{equation}
    H_{2d, C} = \left(DFT_2\right)^{\oplus d} \cdot P \cdot \left(DFT_d\right) ^ {\oplus 2}
    \label{C}
\end{equation}
where $P$ is a permutation which interleaves the outputs of blocks in one layer to the inputs for the next i.e. it routes the $i$-th mode of the $j$-th DFT in the first layer to the $j$th mode of the $i$-th DFT in the second layer.

The optimal choice of $H_d$ depends on the type of success probability which is most useful for the success probability. Table~\ref{probs_table} shows these success probabilities up to $d=5$.

\subsection{Heralded Bell circuit: Version 3}

This circuit can be considered a generalisation of the Bell pair generation circuit from Zou et al.~\cite{Zou2005}. It is also closely related to Version 2c. Reversing the first two layers allows you to swap between the two circuits. The circuit design is shown in Fig.~\ref{bell_state_circuits}c. Success probabilities are summarised in Table~\ref{probs_table}.

\begin{table}[]
\begin{tabular}{l|l|r|r|r|r|r|r}
                                            & \multirow{2}{*}{d} & \multicolumn{1}{c|}{\multirow{2}{*}{v1}} & \multicolumn{3}{c|}{v2} & \multicolumn{1}{c|}{\multirow{2}{*}{v3}} & \multicolumn{1}{c}{\multirow{2}{*}{ZTL}} \\
                      &  &  & \multicolumn{1}{c}{A} &\multicolumn{1}{c}{B} &\multicolumn{1}{c|}{C} &  &                        \\\hline
\multicolumn{1}{l|}{\multirow{4}{*}{$p_a$}} & 2 & 0                      & $6.25 \times 10^{-2}$   & 0.125                   & 0.125                   & 0.125                 & $9.60 \times 10^{-2}$   \\
\multicolumn{1}{l|}{}                       & 3 & $4.06 \times 10^{-4}$  & $1.09 \times 10^{-2}$   & $1.09 \times 10^{-2}$   & $2.19 \times 10^{-2}$   & $2.19 \times 10^{-2}$ & $2.14 \times 10^{-2}$   \\
\multicolumn{1}{l|}{}                       & 4 & $1.65 \times 10^{-5}$  & $1.82 \times 10^{-3}$   & $4.26 \times 10^{-3}$   & $4.35 \times 10^{-3}$   & $4.35 \times 10^{-3}$ & $4.21 \times 10^{-3}$   \\
\multicolumn{1}{l|}{}                       & 5 &                        & $9.66 \times 10^{-5}$   & $9.66 \times 10^{-5}$   & $1.93 \times 10^{-4}$   & $1.93 \times 10^{-4}$ & $7.69 \times 10^{-4}$   \\ \hline
\multicolumn{1}{l|}{\multirow{4}{*}{$p_f$}} & 2 & 0                      & $6.25 \times 10^{-2}$   & 0.1875                   & 0.1875                   & 0.125                & $9.60 \times 10^{-2}$   \\
\multicolumn{1}{l|}{}                       & 3 & $4.06 \times 10^{-4}$  & $1.09 \times 10^{-2}$   & $1.09 \times 10^{-2}$   & $2.19 \times 10^{-2}$   & $2.19 \times 10^{-2}$ & $2.14 \times 10^{-2}$   \\
\multicolumn{1}{l|}{}                       & 4 & $1.65 \times 10^{-5}$  & $1.82 \times 10^{-3}$   & $6.43 \times 10^{-3}$   & $6.43 \times 10^{-3}$   & $4.35 \times 10^{-3}$ & $4.21 \times 10^{-3}$   \\
\multicolumn{1}{l|}{}                       & 5 &                        & $9.66 \times 10^{-5}$   & $9.66 \times 10^{-5}$   & $1.93 \times 10^{-4}$   & $1.93 \times 10^{-4}$ & $7.69 \times 10^{-4}$   \\ \hline
\multicolumn{1}{l|}{\multirow{4}{*}{$p_c$}} & 2 & 0                      & $6.25 \times 10^{-2}$   & 0.1875                  & 0.1875                  & 0.125                 & $9.60 \times 10^{-2}$   \\
\multicolumn{1}{l|}{}                       & 3 & $5.08 \times 10^{-4}$  & $1.09 \times 10^{-2}$   & $1.09 \times 10^{-2}$   & $2.19 \times 10^{-2}$   & $2.19 \times 10^{-2}$ & $2.14 \times 10^{-2}$   \\
\multicolumn{1}{l|}{}                       & 4 & $2.39 \times 10^{-5}$  & $1.82 \times 10^{-3}$   & $6.43 \times 10^{-3}$   & $8.69 \times 10^{-3}$   & $6.60 \times 10^{-3}$ & $4.21 \times 10^{-3}$   \\
\multicolumn{1}{l|}{}                       & 5 &                        & $9.66 \times 10^{-5}$   & $9.66 \times 10^{-5}$   & $1.15 \times 10^{-3}$   & $1.15 \times 10^{-3}$ & $7.69 \times 10^{-4}$  
\end{tabular}

\caption{Success probabilities for heralded qudit Bell state generation using different possible circuits. The labels v1, v2, v3 refer to the Bell state circuits described in sections 1, 2 and 3 of Appendix~\ref{bell_states}. A, B, C refer to the different options for the unitary $H_{2d}$ as described in equations \ref{A}, \ref{B}, \ref{C}. ZTL refers to the zero-transmission-law based Bell pair generators described in Fig.~1b of the main text.
\label{probs_table}}
\end{table}

\section{State generation in the presence of photon distinguishability}
\label{SupplInfo:DistAnalysis}
Here we briefly describe how we simulated the state generation for qudit Bell states of dimensions $2-5$ when using partially distinguishable photons in the heralded high-dimensional entanglement schemes described in the main text. As described below, the best methods to simulate the evolution of partially distinguishable photons in linear-optical interferometers calculate the probabilities for the various output configurations, but do not directly provide their amplitudes. To calculate the full quantum state we therefore use qudit quantum state tomography techniques that allow us to reconstruct the quantum state from the simulated measurement probabilities.

\subsection{Qudit quantum state tomography}
As mentioned in the main text, only simulations of bipartite entangled states with $N=2$ remain tractable on a standard laptop. Therefore we here limit discussion to quantum state tomography of systems of two qudits. In general the density matrix for such a state may be written as~\cite{Caves2000,Thew2002}

\begin{equation}
\label{eqn:twoQuditState}
    \rho=\frac{1}{d^2}\sum_{i,j=0}^{d^2-1}r_{i,j}G_{i}\otimes G_{j},
\end{equation}
where $G_{i}$ are the generators of $SU(d)$. These form an orthogonal operator basis in the space of linear operators for a qudit Hilbert space and obey the normalisation $\mathrm{Tr}(G_{i}G_{j})=2\delta_{ij}$. There are $d^2-1$ generators, which are the Pauli matrices for for $d=2$ and the Gell-Mann matrices for $d\geq3$. We also include the rescaled identity $G_{0}=\sqrt{2/d}\times\mathbb{I}$ for a total of $d^2$ Hermitian operators with which to decompose a qudit. The coefficients $r_{i,j}$ can be calculated using the expectations of these generators by

\begin{equation}
\label{eqn:twoQuditStateCoefficients}
    r_{i,j}=\left(\frac{d}{2}\right)^2 \times \langle G_{i}\otimes G_{j}\rangle = \left(\frac{d}{2}\right)^2\times\mathrm{Tr}\left[\rho(G_{i}\otimes G_{j})\right].
\end{equation}

\subsection{Output probabilities with  partial distinguishability}
Given the generation circuit in Fig.~1b for qudit Bell states, we need to calculate the expectations of these operators $G_{i}$ for different values of photon indistinguishability $\vert\langle\psi_{i}\vert\psi_{j}\rangle\vert^2$ as mentioned in the main text. To do so, we will need to find unitaries that rotate between the operator eigenbases and the qudit basis, since the latter is the one in which we detect photons.

We first take the spectral decomposition of the Hermitian operator

\begin{equation}
    G_{i}=\sum_{j=0}^{d-1}\lambda_{j}^{(i)}\vert v_{j}^{(i)}\rangle\langle v_{j}^{(i)}\vert,
\end{equation}
where $\{\vert v_{j}^{(i)}\rangle\}$ are the eigenvectors and $\lambda_{j}^{(i)}$ are the associated eigenvalues. In a similar way, the tensor product of a pair of these operators can be written as

\begin{equation}
    G_{i}\otimes G_{j} = \sum_{k,l=0}^{d-1}\lambda_{k}^{(i)}\lambda_{l}^{(j)}\left(\vert v_{k}^{(i)}\rangle\otimes\vert v_{l}^{(j)}\rangle\right)\left(\langle v_{k}^{(i)}\vert\otimes\langle v_{l}^{(j)}\vert\right).
\end{equation}
Next we define the unitary

\begin{equation}
    U_{i}=\sum_{j=0}^{d-1}\vert j\rangle\langle v_{j}^{(i)}\vert
\end{equation}
that performs a rotation from the eigenvector basis to the qudit basis $\{\vert j\rangle\}$. Given the heralded state $\rho$ from the circuit in Fig.~1b defined over the modes for two qudits, we now consider appending unitary $U_{i}$ to the modes defining the first qudit and $U_{j}$ to those defining the second qudit. The probability of detecting the first photon in mode $k$ and the second in mode $l$ is then

\begin{equation}
    P(k,l\vert i,j)=\mathrm{Tr}\left[\rho\left(\vert v_{k}^{(i)}\rangle\otimes\vert v_{l}^{(j)}\rangle\right)\left(\langle v_{k}^{(i)}\vert\otimes\langle v_{l}^{(j)}\vert\right)\right].
\end{equation}
Hence if we calculate these probabilities, given a successful heralding pattern, then the expectation value is given by $\langle G_{i}\otimes G_{j}\rangle = \sum_{k,l}\lambda_{k}^{(i)}\lambda_{l}^{(j)}\times P(k,l\vert i,j)$.

To calculate these probabilities we use the method by Tichy~\cite{Tichy2015}. This allows calculation of the probability of some input configuration of partially distinguishable photons to propagate through a linear circuit and end up in a specified output configuration. The input configuration is denoted by the mode occupation list $\vec{r}=(r_{1},...,r_{m})$, where $r_{j}$ indicates the number of photons in input mode $j$. For the qudit Bell state generation circuit in Fig.~1b, all modes are occupied by a single photon so all $r_{j}=1$ and $\sum_{j}r_{j}=m=2d+1$. The output configuration can be labelled by $\vec{s}$ and $\sum_{j}s_{j}=2d+1$, and here the heralding condition means $s_{0}=2d-1$. The probability is given by

\begin{equation}
\label{eqn:scatteringProbs}
    P(\vec{r},\vec{s},U)=\frac{1}{\prod_{j}r_{j}!s_{j}!}\sum_{\sigma,\rho\in S_{m}}\prod_{j=1}^{m}\left(M_{\sigma_{j},j}M^{*}_{\rho_{j},j}\mathcal{S}_{\rho_{j},\sigma_{j}}\right).
\end{equation}
$U$ is the overall unitary circuit that includes the DFT for generation, and also the unitaries for basis rotation $U_{i}$ and $U_{j}$ on the modes defining the first and second qudits respectively. $M$ is the effective scattering matrix obtained by selecting rows and columns from $U$ with multiplicities given by the occupation numbers of $\vec{r}$ and $\vec{s}$ respectively. For example, $\vec{r}=(1,1)$ and $\vec{s}=(2,0)$ would correspond to taking each row once and the first column twice to construct $M$. $\mathcal{S}$ is obtained by taking rows and columns, with multiplicities given by the input occupation numbers, from the Hermitian matrix of pairwise overlaps $\mathcal{S}_{ij}=\langle\psi_{i}\vert\psi_{j}\rangle$. We assume that all photons have the same pairwise overlap. $\rho$ and $\sigma$ are elements of the permutation group $S_{m}$. $M^{*}_{\rho_{j},j}$ is the element-wise complex conjugate of $M$ with its rows permuted according to $\rho$. There are $n!^2$ terms in this double sum over the permutation group. However Tichy provides an adaption of Ryser's algorithm to reduce this to a sum over $2^{2n}$ terms instead, making our simulations considerably more tractable~\cite{Tichy2015}.

For each qudit Bell state dimension $d$, we choose a set of pairwise overlaps and for each we calculate the scattering probabilities when appending the different unitaries $U_{i}$ and $U_{j}$. After correcting for the marginal heralding probability, we then calculate the expectations $\langle G_{i}\otimes G_{j}\rangle$. Finally we use linear inversion in Eq.~\ref{eqn:twoQuditState} to reconstruct the heralded state $\rho$. We are effectively postselecting on it being in the computational space, and in fact as photon distinguishability increases, the state moves outside the computational space and into the full Fock space of 2 photons in $2d$ modes.

\subsection{Fidelity and logarithmic negativity}
Given the reconstructed state $\rho$, we then want to assess our scheme's robustness to photon distinguishability.
We do this by calculating the fidelity with the ideal qudit Bell state and the logarithmic negativity.
The fidelity of the reconstructed state $\rho$ with the ideal state $\sigma$ is given by $F=\left(\mathrm{Tr}\sqrt{\sqrt{\sigma}\rho\sqrt{\sigma}}\right)^{2}$.
The logarithmic negativity is given by $E_{\mathcal{N}}(\rho)=\mathrm{log}_{2}\vert\vert \rho^{T_{A}}\vert\vert_{1}$, where $\rho^{T_{A}}$ is the partial transpose of $\rho$ with respect to party $A$ of the bipartite state, and the trace norm $\vert\vert X\vert\vert_{1} = \mathrm{Tr}\sqrt{X^{\dagger}X}$.
In Fig.~4 of the main text we plot these quantities for the states generated in our scheme when using photons of different indistinguishabilities.

\end{document}